\documentclass[12pt]{article}
\usepackage{amsmath,bm,amsfonts,color,amsthm, amssymb}
\usepackage[right=1.0in, left=1.0in, top=1.2in, bottom=2.0in]{geometry}
\usepackage{graphicx}
\usepackage{natbib}
\usepackage{xcolor}
\usepackage{enumitem}
\usepackage{subcaption,siunitx,booktabs}
\usepackage[citecolor=blue,urlcolor=blue,allcolors=blue]{hyperref}
\usepackage{arydshln}

\usepackage{titlesec}
\titlelabel{\thetitle.\quad}

\usepackage{tikz}
\usepackage{transparent}

\usetikzlibrary{positioning}
\usetikzlibrary{shadows}
\usepgflibrary{shapes}

\usepackage{mathtools}

\usepackage{apptools}
\AtAppendix{\counterwithin{lemma}{section}}

\AtAppendix{\counterwithin{proposition}{section}}

\usepackage{titlesec}

\titleformat*{\section}{\large\bfseries}
\titleformat*{\subsection}{\large\bfseries}
\titleformat*{\subsubsection}{\large\bfseries}
\titleformat*{\paragraph}{\large\bfseries}
\titleformat*{\subparagraph}{\large\bfseries}

\usepackage{titlesec}
\titlelabel{\thetitle\quad}

\theoremstyle{definition}

\numberwithin{mytheorem}{section}
\usepackage{setspace}

\newtheorem{myremark}{Remark}

\numberwithin{myassumption}{section}

\numberwithin{definition}{section}

\title{An application of a small area procedure with correlation between measurement error and sampling error to the Conservation Effects Assessment Project}

  \author{Emily Berg\thanks{
    contact the corresponding author at \href{mailto:emilyb@iastate.edu}{emilyb@iastate.edu}}\hspace{.2cm}\\
    Iowa State University\\
    and \\
    Sepideh Mosaferi \\
    University of Massachusetts Amherst}

\begin{document}

\maketitle
\begin{abstract}

County level estimates of mean sheet and rill erosion from the Conservation Effects Assessment Project (CEAP) are useful for program development and evaluation. Since county sample sizes in the CEAP survey are insufficient to support reliable direct estimators, small area estimation procedures are needed. The quantity of water runoff is a useful covariate but is unavailable for the full population. We use  an estimate of mean runoff from the CEAP survey as a covariate in a small area model with sheet and rill erosion as the response.  As the runoff and sheet and rill erosion are estimators from the same survey, the measurement error in the covariate is important as is the correlation between the measurement error and the sampling error.  We conduct a detailed investigation of small area estimation in the presence of a correlation between the measurement error in the covariate and the sampling error in the response.  In simulations, the proposed predictor is superior to small area predictors that assume the response and covariate are uncorrelated or that ignore the measurement error entirely.  

\end{abstract}

\section{Introduction}  \label{sec:Intro}

The Conservation Effects Assessment Project (CEAP) is a program comprised of several surveys that are intended to evaluate the environmental impacts of agricultural production. We consider data from a CEAP survey of cropland that was conducted over the period 2003-2006.  An important variable collected in CEAP is sheet and rill erosion (soil loss due to the flow of water). County estimates of sheet and rill erosion can improve the efficiency of allocation of resources for conservation efforts. Sample sizes in CEAP are too small to support reliable direct county estimates.  Past analyses have explored a variety of issues that arise in the context of small area estimation using CEAP data \citep{erciulescu2016small, berg2014small, lyu2020empirical, berg2019prediction}.  

Traditional small area estimation procedures utilize population-level auxiliary information from censuses or administrative databases  \citep{rao2015small, pfeffermann2013new, jiang2006mixed}. A critical assumption underlying the seminal \cite{fay1979estimates} predictor is that one can condition on the observed value of the covariate. As discussed in \cite{lyu2020empirical}, the task of obtaining covariates that are related to sheet and rill erosion and are known for the full population of cropland of interest is difficult. Use of variables collected in the CEAP survey as covariates is therefore desirable. We use an estimate of mean water runoff from the CEAP survey as a covariate in an area level model with sheet and rill erosion as the response. As the covariate and response are both estimates from the CEAP survey, the analysis should recognize not only the sampling error in the covariate but also the correlation between the   covariate and the  response. 

If the covariate is an estimator from a sample survey, naive application of standard Fay-Herriot procedures can lead to erroneous inferences \citep{bell2019measurement, arima2017multivariate}. A widely used technique to model sampling error in the covariate is to employ either a structural \citep{ghosh2006empirical, Arima2012, torabi2009empirical, torabi2012small} or a functional \citep{ghosh2007empirical, datta2010pseudo, torabi2011small}  measurement error model. In the structural model, the latent covariate is stochastic, while the functional model treats the unobserved covariate as fixed \citep{fuller2009measurement, carroll2006measurement}. 

\cite{ybarra2008small} develops predictors for an area level model in which a covariate is subject to functional measurement error. \cite{arima2015bayesian}, \cite{arima2017multivariate}, and \cite{burgard2021small} extend \cite{ybarra2008small} to bivariate and Bayesian frameworks. \cite{burgard2020fay} conducts an analysis of the model of \cite{ybarra2008small} under the stronger assumption that the random terms have normal distributions.   \cite{bell2019measurement} compares the properties of functional and structural measurement error models to the naive Fay-Herriot model. \cite{mosaferi2021transformed} extends \cite{ybarra2008small} to lognormal data.  All of these works assume that the measurement error in the covariate and the sampling error in the response are uncorrelated. 
 
 Models in which the measurement error and the sampling error are correlated have received little attention in the small area estimation literature. \cite{franco2019using} defines a bivariate model that is equivalent to a structural model with a correlation between the measurement error and the sampling error. We adopt the functional modeling approach, which, unlike the structural model, requires no assumptions about the distribution of the latent covariate.  \cite{kim2015small} permits a correlation between the covariate and response but  conceptualizes the parameter of interest as the unobserved value of a covariate that is measured with error.  \cite{ybarra2003} generalizes the functional measurement error model of \cite{ybarra2008small} to allow for a correlation between the measurement error and the sampling error. \cite{burgard2022measurement} uses likelihood-based arguments to derive a predictor for a model in which the measurement error and the sampling error are correlated. 
 
We conduct a thorough analysis of a model in which the measurement error and the sampling error are correlated. Our work expands on \cite{ybarra2003} in several dimensions. We conduct extensive simulation studies in a framework where the measurement error in the covariate is correlated with the sampling error in the response. We rigorously discuss  the theoretical properties of the predictors with estimated parameters. Further, we provide comprehensive software at \url{https://github.com/emilyjb/SAE-Correlated-Errors/}.

The rest of this manuscript is organized as follows. We derive a predictor using properties of the bivariate normal distribution in Section 2.1. We propose estimators of the fixed model parameters in Section 2.2, and we study the theoretical properties of the proposed estimators. Then, we derive the mean squared prediction error (MSPE) of the proposed predictor. We also conduct extensive simulations to assess the properties of the proposed procedures in Section 3. We apply the proposed method to data from the CEAP survey in Section 4.  We conclude in Section 5.

\section{ Model and Predictor}

We define an area-level model, where the measurement error in the covariate is correlated with the sampling error in the response.  We denote the true, unknown value of the covariate by $\bm{x}_{i} = (x_{i1}, \ldots, x_{ip})'$ for $i = 1,\ldots, n$, where $n$ is the total number of small areas. We do not observe $\bm{x}_{i}$ directly. Instead, we observe a contaminated version of $\bm{x}_{i}$ denoted by $\bm{W}_{i}$. In the common situation, $\bm{W}_{i}$ represents an estimator of $\bm{x}_{i}$ obtained from a survey. 
The measurement error is functional instead of structural because $\bm{x}_{i}$ is regarded as a fixed quantity. 

The parameter of interest is   
\begin{align*}
\theta_{i} = \beta_{0}+ \bm{x}_{i}'\bm{\beta}_{1} + b_{i}, 
\end{align*}
where $b_{i}\stackrel{iid}{\sim} N(0,\sigma^{2}_{b} )$ for $i = 1,\ldots, n$. We let $Y_{i}$ denote an estimator of $\theta_{i}$. The variables representing observable quantities are then $\{ ( \bm{W}_{i}', Y_{i})': i = 1,\ldots, n\}$. We define a model for $(\bm{W}_{i}', Y_{i})'$ as, 
\begin{align}\label{mecor}
Y_{i} & = \beta_{0} + \bm{x}_{i}'\bm{\beta}_{1} + b_{i} + e_{i},  \\  \nonumber 
\bm{W}_{i} & = \bm{x}_{i} + \bm{u}_{i}, 
\end{align}
and $(\bm{u}_{i}', e_{i})' \stackrel{ind}{\sim} MVN(\bm{0}, \bm{\Psi}_{i}),$ where $(\bm{u}_{i}', e_{i})$ is independent of $b_{i}$ for $i = 1,\ldots, n$. We assume that $\bm{\Psi}_{i}$ is known.

Typically, $\bm{\Psi}_{i}$ is the design-variance of $(\bm{W}_{i}', Y_{i})'$, and $\bm{\Psi}_{i}$ is constructed from the unit-level data using standard procedures for complex surveys.   We parametrize $\bm{\Psi}_{i}$ as 
\begin{align*}
\bm{\Psi}_{i} = \left( \begin{array}{cc}  \bm{\Psi}_{uui} & \bm{\Psi}_{uei} \\ \bm{\Psi}_{uei}' & \psi_{eei} \end{array}\right). 
\end{align*}
The component $\bm{\Psi}_{uei}$ captures the correlation between the measurement error and the sampling error. We denote the fixed vector of parameters which needs to be estimated as $\bm{\omega} = (\beta_{0}, \bm{\beta}_{1}', \sigma^{2}_{b})'$. The objective is to predict the small area parameter $\theta_{i}$.

\begin{myremark}
The model in (\ref{mecor}) has strong connections to other models in the small area estimation literature. 
The model is identical to the model of \cite{ybarra2003}. If $\bm{\Psi}_{uei} = \bm{0}$, then the model in (\ref{mecor}) simplifies to the functional measurement error models of \cite{ybarra2008small} and \cite{burgard2020fay}.   The structure of our model is also similar to the one given by \cite{kim2015small}. The parameter of interest in our framework is the conditional mean of the response denoted by $\theta_{i}$, which differs from the parameter of interest in \cite{kim2015small}, the unobserved covariate, $\bm{x}_{i}$.    
 In small area estimation, the parameter of interest is usually the mean of a response variable. Thus, we think that our formulation is more useful to practitioners than that of \cite{kim2015small}.
\end{myremark}

\begin{myremark}
In many situations, as in the CEAP data analysis of Section 4, the observed covariate is an estimator from a sample survey. In this case, the error in the observed covariate is a sampling error.   Because the model (\ref{mecor}) has the form of a measurement error model, we refer to the random term $\bm{u}_{i}$ as a measurement error instead of a sampling error. This terminology is common in the small area estimation literature \citep{ybarra2008small}. 
\end{myremark}

\subsection{ Predictor as a Function of the True Model Parameters}

We first define a predictor as a function of the unknown $\bm{\omega}$. This predictor is also defined in \cite{ybarra2003}. We provide a derivation that differs slightly from that of \cite{ybarra2003}. One can express the parameter of interest as  $\theta_{i} = Y_{i} - e_{i}$. We   define a predictor of the parameter of interest as $Y_{i} - \hat{e}_{i}$, where $\hat{e}_{i}$ is an appropriately defined predictor of $e_{i}$. We now proceed to develop a form for $\hat{e}_{i}$. As in \cite{fuller2009measurement} and \cite{ybarra2008small}, define 
\begin{align*}
v_{i} = Y_{i} - \beta_{0} - \bm{\beta}_{1}'\bm{W}_{i} = b_{i} + e_{i} - \bm{\beta}_{1}'\bm{u}_{i}. 
\end{align*}
Then, using properties of the bivariate normal distribution (as explained in Appendix A),  
\begin{align}\label{condnormei}
e_{i} \mid v_{i} \sim N\left( \frac{ \psi_{eei} - \bm{\beta}_{1}'\bm{\Psi}_{uei} }{ \sigma^{2}_{b} + \sigma^{2}_{\delta i } } v_{i},\psi_{eei} -  \frac{(\psi_{eei} - \bm{\beta}_{1}'\bm{\Psi}_{uei})^{2}}{ \sigma^{2}_{b} + \sigma^{2}_{\delta i}  }\right), 
\end{align}
where  $\sigma^{2}_{\delta i }= \bm{\beta}_{1}'\bm{\Psi}_{uui}\bm{\beta}_{1}+ \psi_{eei} -2\bm{\beta}_{1}'\bm{\Psi}_{uei}$. A predictor of $\theta_{i}$ is then 
\begin{align}\label{thetatildei}
\tilde{\theta}_{i} = \tilde{\theta}_{i}(\bm{\omega}) = Y_{i} - \hat{e}_{i}(\bm{\omega}),
\end{align}
where 
\begin{align*}
\hat{e}_{i}(\bm{\omega}) =  E(e_i|v_i)= \frac{ \psi_{eei} - \bm{\beta}_{1}'\bm{\Psi}_{uei} }{ \sigma^{2}_{b} + \sigma^{2}_{\delta i} } v_{i}.
\end{align*}
The MSPE of $\tilde{\theta}_{i}(\bm{\omega})$ is 
\begin{align}\label{m1iomega} 
MSPE(\tilde{\theta}_{i}(\bm{\omega})) & = E[(Y_{i} - e_{i} - (Y_{i} - \hat{e}_{i}(\bm{\omega})) )^{2}] \\ \nonumber 
& =  V(e_{i} \mid v_{i})   \\ \nonumber
& = \psi_{eei} -  \frac{(\psi_{eei} - \bm{\beta}_{1}'\bm{\Psi}_{uei})^{2}}{ \sigma^{2}_{b} + \sigma^{2}_{\delta i}  } := M_{1i}(\bm{\omega}). 
\end{align}
In (\ref{m1iomega}), $V(e_{i} \mid v_{i}) = E[V(e_{i} \mid v_{i})]$ because $V(e_{i} \mid v_{i})$ does not depend on any random variables. 

\begin{myremark}
We use the properties of the bivariate normal distribution to derive the predictor (\ref{thetatildei}).  A different way to develop a predictor is to find the convex combination of $Y_{i}$ and $\beta_{0}+ \bm{\beta}_{1}'\bm{W}_{i}$ that minimizes the MSPE. \cite{ybarra2003} demonstrates that the predictor (\ref{thetatildei}) is the minimum MSPE convex combination of $Y_{i}$ and $\beta_{0}+ \bm{\beta}_{1}'\bm{W}_{i}$ under moment conditions that do not require normality.
\end{myremark}

\begin{myremark}
\cite{burgard2022measurement} define a predictor for a generalization of the model (\ref{mecor}). The predictor proposed in (\ref{thetatildei}) differs from that of \cite{burgard2022measurement}. In the supplement, we provide empirical evidence that the predictor (\ref{thetatildei}) is more efficient than the predictor of \cite{burgard2022measurement}. 
\end{myremark}

\subsection{Estimation of Parameters}

We require estimators of $\beta_{0}$, $\bm{\beta}_{1}$, and  $\sigma^{2}_{b}$.  We estimate the regression coefficients by matching the sample moments with their theoretical expectations. Define  $\bm{\zeta}_{1}  = n^{-1}\sum_{i =1}^{n} Y_{i}\bm{W}_{i} - n^{-1}\sum_{i = 1}^{n} \bm{\Psi}_{ue i},$ $\zeta_{2}  = n^{-1}\sum_{i = 1}^{n}Y_{i}, $ $
\bm{\zeta}_{3}  = n^{-1}\sum_{i=1}^{n} \bm{W}_{i},$ and $\bm{\zeta}_{4}  = n^{-1}\sum_{i =1}^{n} \bm{W}_{i}\bm{W}_{i}' -  n^{-1}\sum_{i = 1}^{n} \bm{\Psi}_{uui}.$  
Then,  $E[\zeta_{2}] = \beta_{0} + \bm{\beta}_{1}'E[\bm{\zeta}_{3}]$ and $E[\bm{\zeta}_{1}] = \beta_{0}E[\bm{\zeta}_{3}]+ E[\bm{\zeta}_{4}]\bm{\beta}_{1}.$ We define an estimator of $(\beta_{0}, \bm{\beta}_{1})'$ as 
\begin{align*}
\left( \begin{array}{c} \hat{\beta}_{0} \\ \hat{\bm{\beta}}_{1}\end{array}\right) & = \left( \begin{array}{cc} 1 & \bm{\zeta}_{3}' \\
\bm{\zeta}_{3} &  \bm{\zeta}_{4} \end{array}\right)^{-1}\left( \begin{array}{c} \zeta_{2} \\ \bm{\zeta}_{1}\end{array}\right). 
\end{align*}
Theorem 1 of  Appendix B states that the estimator of the regression coefficients is consistent, where we simplify and only consider a univariate covariate.  We outline the proof of Theorem 1 in Appendix B and provide further details in the supplementary material. 

The estimator of $\sigma^{2}_{b}$ defined in \cite{ybarra2008small} can be easily adjusted to the case of correlated errors, as in \cite{ybarra2003}.  Following \cite{ybarra2008small}, one can define an estimator of $\sigma^{2}_{b}$ as 
\begin{align}\label{sig2b0} 
\hat{\sigma}^{2}_{b, YL} = \frac{1}{n-p-1}\sum_{i =1}^{n}\{ (Y_{i} - \hat{\beta}_{0} - \hat{\bm{\beta}}_{1}'\bm{W}_{i} )^{2} - \psi_{eei} - \hat{\bm{\beta}}_{1}'\bm{\Psi}_{uui}\hat{\bm{\beta}}_{1} + 2\hat{\bm{\beta}}_{1}'\bm{\Psi}_{uei}\}.
\end{align} 
A drawback of the estimator (\ref{sig2b0}) is that it can be negative.  Thus, we use a profile likelihood.

We define the profile likelihood for estimating $\sigma^{2}_{b}$ by 
\begin{align}\label{ladj}
L(\sigma^{2}_{b}\mid \hat{\beta}_{0}, \hat{\bm{\beta}}_{1}) =  \left[\prod_{i = 1}^{n}\left( \frac{1}{\sqrt{2\pi (\sigma^{2}_{b} + \hat{\sigma}^{2}_{\delta i}) }}\right)\right]\mbox{exp}\left( - 0.5\sum_{i=1}^{n}\frac{(Y_{i} - \hat{\beta}_{0} - \hat{\bm{\beta}}_{1}'\bm{W}_{i})^{2}}{ \sigma^{2}_{b} + \hat{\sigma}^{2}_{\delta i} }\right),
\end{align} 
where $\hat{\sigma}^{2}_{\delta i} = \psi_{eei} + \hat{\bm{\beta}}_{1}'\bm{\Psi}_{uui}\hat{\bm{\beta}}_{1} - 2\hat{\bm{\beta}}_{1}'\bm{\Psi}_{uei}$. The estimator of unknown parameter $\sigma^2_b$ is
\begin{align}\label{sig2hatb}
\hat{\sigma}^{2}_{b} = \mbox{argmax}_{\sigma^{2}_{b} \geq 0} L(\sigma^{2}_{b} \mid \hat{\beta}_{0}, \hat{\bm{\beta}}_{1}),
\end{align}
where the maximization is over the parameter space for $\sigma^{2}_{b}$. The profile likelihood estimator is similar to a maximum likelihood estimator (MLE) in the sense that the estimator does not account for the loss of degrees of freedom from estimating regression coefficients.  Another possibility is to construct a restricted ML (REML)-type estimator to improve upon the properties of $\hat{\sigma}^{2}_{b}$, and this is a possible future research direction. We give theoretical consideration to the properties of $\hat{\sigma}^{2}_{b}$ for a univariate covariate.  Theorem 2 in Appendix B states that $\hat{\sigma}^{2}_{b}$ is a consistent estimator of $\sigma^{2}_{b}$. We present a proof of Theorem 2 in the supplement.

\begin{myremark}
For the estimation of $\sigma^{2}_{b}$, \cite{li2010adjusted} proposed an adjusted likelihood. We tried this technique and found that the resulting estimator can have a positive bias in simulations. We therefore use the profile likelihood function  (\ref{ladj}) for our simulations and data analysis. The benefits of our proposed estimator are that the estimator is simple and has tractable theoretical properties, as we discuss in Appendix B. 
\end{myremark}

\begin{myremark}
It may be observed that we use a likelihood-based estimator of $\sigma^{2}_{b}$ and a moment-based estimator of the regression coefficients. An alternative is to develop a likelihood-based estimator of the regression coefficients, along the lines of \cite{burgard2020fay}. We prefer the moment-based estimator for two main reasons. First, the estimator can be calculated in one step, enabling a computationally simple procedure. Second, the moment-based estimator is robust to the assumption of normality, as we demonstrate through the simulation study. We prefer the profile-likelihood estimator of $\sigma^{2}_{b}$ over the moment-based estimator for the purpose of avoiding negative estimates. 
\end{myremark}

\subsection{Predictors with Estimated Parameters}

We evaluate the predictor (\ref{thetatildei}) at the estimator of $\bm{\omega}$. We define the predictor as 
\begin{align}\label{thetahati}
\hat{\theta}_{i} = \tilde{\theta}_{i}(\hat{\bm{\omega}}) = Y_{i}  - \hat{e}_{i}(\hat{\bm{\omega}}),  
\end{align}

\noindent where $\hat{\bm{\omega}} = (\hat{\beta}_{0}, \hat{\bm{\beta}}_{1}', \hat{\sigma}^{2}_{b})'$. The vector of estimated parameters $\hat{\bm{\omega}}$ is obtained using the procedure of Section 2.2. 
The MSPE of $\hat{\theta}_{i}$ decomposes into a sum of three terms as 

\begin{align*}
MSPE(\hat{\theta}_{i})  = & E[(Y_{i} - \hat{e}_{i}(\hat{\bm{\omega}}) -  (Y_{i} - e_{i})  )^{2}] \\ \nonumber 
  = & E[(\hat{e}_{i}(\hat{\bm{\omega}}) - e_{i}  )^{2}]  \\ \nonumber 
  = & E[(\hat{e}_{i}(\bm{\omega}) - e_{i})^{2}] + E[(\hat{e}_{i}(\hat{\bm{\omega}}) - \hat{e}_{i}(\bm{\omega}))^{2}] \\ \nonumber
  + & 2E[(\hat{e}_{i}(\hat{\bm{\omega}}) - \hat{e}_{i}(\bm{\omega}) )(\hat{e}_{i}( \bm{\omega}) - e_{i}  )] \\
  = & V(e_{i} \mid v_{i})  + E[(\hat{e}_{i}(\hat{\bm{\omega}}) - \hat{e}_{i}(\bm{\omega}) )^{2}] + 2E[(\hat{e}_{i}(\hat{\bm{\omega}}) - \hat{e}_{i}(\bm{\omega}) )(\hat{e}_{i}( \bm{\omega}) - e_{i}  )]\\ \nonumber 
  = & M_{1i}(\bm{\omega})  + M_{2i} + 2E[(\hat{e}_{i}(\hat{\bm{\omega}}) - \hat{e}_{i}(\bm{\omega}) )(\hat{e}_{i}( \bm{\omega}) - e_{i}  )]. 
\end{align*}

The first term, $M_{1i}(\bm{\omega}) = V(e_{i} \mid v_{i}) = E[V(e_{i} \mid v_{i}) ]$, is defined in (\ref{m1iomega}). The second term, $M_{2i} = E[(\hat{e}_{i}(\hat{\bm{\omega}}) - \hat{e}_{i}(\bm{\omega}) )^{2}]$,  accounts for the variance of $\hat{\bm{\omega}}$. Consider the cross term defined as $2E[(\hat{e}_{i}(\hat{\bm{\omega}}) - \hat{e}_{i}(\bm{\omega}) )(\hat{e}_{i}( \bm{\omega}) - e_{i}  )]$. Suppose $\hat{\bm{\omega}}$ is independent of $e_{i}$ given $v_{i}$. Then, 
\begin{align*}
E[(\hat{e}_{i}(\hat{\bm{\omega}}) - \hat{e}_{i}(\bm{\omega}) )(\hat{e}_{i}( \bm{\omega}) - e_{i}  )]  &= E[E[ (\hat{e}_{i}(\hat{\bm{\omega}}) - \hat{e}_{i}(\bm{\omega}) )(\hat{e}_{i}( \bm{\omega}) - e_{i}  )  \mid v_{i}, \hat{\bm{\omega}}]] \\ \nonumber 
& =  E[(\hat{e}_{i}(\hat{\bm{\omega}}) - \hat{e}_{i}(\bm{\omega}) )E[ ( \hat{e}_{i}(\bm{\omega}) - e_{i}  )  \mid v_{i}, \hat{\bm{\omega}}]] \\ \nonumber 
& =  E[(\hat{e}_{i}(\hat{\bm{\omega}}) - \hat{e}_{i}(\bm{\omega}) )E[ (\hat{e}_{i}(\bm{\omega}) - e_{i}  )  \mid v_{i}]] = 0. 
\end{align*}
The final equality holds because $\hat{e}_{i}(\bm{\omega}) = E[e_{i}\mid v_{i}]$, and the preceding equality holds by the assumption that $\hat{\bm{\omega}}$ is independent of $e_{i}$ given $v_{i}$. The MSPE of $\hat{\theta}_{i}$ then decomposes into a sum of two terms as 
\begin{align*}
MSPE(\hat{\theta}_{i}) =  M_{1i}(\bm{\omega})  + M_{2i}. 
\end{align*}

We use a plug-in estimator of $M_{1i}(\bm{\omega})$ defined as 
\begin{align*}
\hat{M}_{1i} = M_{1i}(\hat{\bm{\omega}}),  
\end{align*}
where $M_{1i}(\bm{\omega})$ is defined in (\ref{m1iomega}). We use the jackknife technique to estimate $M_{2i}$ as well as the bias of $\hat{M}_{1i}$ for $M_{1i}(\bm{\omega})$. Let $\hat{\bm{\omega}}^{(k)}$ denote the estimator of $\bm{\omega}$ with area $k$ omitted. The jackknife estimator of $M_{2i}$ is defined as
\begin{align*}
\hat{M}_{2i,JK} =  \sum_{k = 1}^{n}(\hat{e}_{i}(\hat{\bm{\omega}}^{(k)}) - n^{-1}\sum_{k = 1}^{n}\hat{e}_{i}(\hat{\bm{\omega}}^{(k)})    )^{2}.
\end{align*}
The jackknife estimate of the bias of the estimator of $M_{1i}$ is 
\begin{align*}
\hat{b}_{i,JK} = n^{-1}\sum_{k = 1}^{n}M_{1i}(\hat{\bm{\omega}}^{(k)}  ) - \hat{M}_{1i}(\hat{\bm{\omega}}). 
\end{align*}
The estimator of the MSPE is then defined as 
\begin{align}\label{msepred} 
\hat{MSPE}_{i} =  \hat{M}_{1i} +\hat{M}_{2i,JK}  - \hat{b}_{i,JK}.  				
\end{align} 

\begin{myremark}
 The simplifying assumption that $e_{i}$ is independent of $\hat{\bm{\omega}}$ given $v_{i}$ facilitates construction of a simple MSPE estimator. The simulation studies presented in Section 3 verify that the MSPE estimator, constructed under this assumption, has good properties. 
\end{myremark} 

\begin{myremark}
Alternatives to the jackknife variance estimator are Taylor linearization and the bootstrap. For this model, Taylor linearization is possible, but the operations are tedious. We prefer the jackknife relative to Taylor linearization for simplicity of implementation. 
\end{myremark}

\begin{myremark}
We use the assumption of normality when formulating the predictor and when proving that the parameter estimators are consistent. The procedures, however, do not rely heavily on the normality assumption. The development of the predictor in \cite{ybarra2003} as the optimal convex combination between the direct estimator and $\beta_{0} + \bm{\beta}_{1}'\bm{W}_{i}$ does not require normality. The estimators of regression coefficients remain consistent under suitable assumptions on the fourth moments. We study the robustness of the prediction procedure to departures from normality through simulations.  
\end{myremark}

\section{ Simulations}

We conduct simulations with two goals. The first is to understand the effect of the nature of $\bm{\Psi}_{i}$ on the properties of the predictor. The second is to assess effects of departures from normality. We simulate data from normal distributions, as specified in model (\ref{mecor}). For the simulations of Section 3.1, we use unequal $\bm{\Psi}_{i}$. For the simulations of Section 3.2, we use a constant value of $\bm{\Psi}_{i} = \bm{\Psi}$ for $i = 1,\ldots, n$.  We generate data from $t$ distributions in Section 3.3. 
For both normal and $t$ distributions, we  use a univariate covariate so that $\bm{\Psi}_{i}$ is a 2$\times$2 matrix, and $\beta_{1}$ is a scalar. We generate a fixed set of $x_{i}$ as independent chi-square random variables with 5 degrees of freedom.  We set $(\beta_{0}, \beta_{1}, \sigma^{2}_{b}) = (1, 2, 0.36)$. 

As one of the simulation objectives is to understand the effects of the form of $\bm{\Psi}_{i}$ on the properties of the predictor, we define a general form for $\bm{\Psi}_{i}$ by 
\begin{align}\label{unequalpsi} 
\bm{\Psi}_{j} &  = (0.75 + 0.25 j)^{2}\mbox{diag}( \sqrt{a}, \sqrt{b}  )\left( \begin{array}{cc} 1& \rho \\ \rho &  1 \end{array}\right)\mbox{diag}( \sqrt{a}, \sqrt{b}  ), 
\end{align}
for $j = 1,2,3,4$. We conduct simulations with equal and uneuqal $\bm{\Psi}_{i}$. For the simulations with unequal $\bm{\Psi}_{i}$, we set one-fourth of the $\{\bm{\Psi}_{i}: i = 1,\ldots, n\}$ equal to $\bm{\Psi}_{j}$ for $j = 1,2,3,4$. Areas $(j-1)n/4 + 1,\ldots, (jn)/4$ are assigned $\bm{\Psi}_{j}$.  For the simulations with equal $\bm{\Psi}_{i}$, we set $\bm{\Psi}_{i} = \bm{\Psi}_{1}$ for $i = 1,\ldots, n$.  The configurations are chosen to reflect a range of conditions.  

We define 8 simulation configurations by four combinations of $(a,b,\rho)$ and two sample sizes. First, we set $(a, b, \rho) = (0.25, 0.75, 0.2)$. For this configuration, the measurement error variance is smaller than the sampling error variance, and the correlation between the measurement error and the sampling error is 0.2. We then increase the correlation and set $(a, b, \rho) = (0.25, 0.75, 0.8)$. Next, we reverse $a$ and $b$ so that the measurement error variance exceeds the sampling error variance. For the third and fourth choices of $\bm{\Psi}_{i}$, we define  $(a, b, \rho) = (0.75, 0.25, 0.2)$ and  $(a, b, \rho) = (0.75, 0.25, 0.8)$. For each choice of $\bm{\Psi}_{i}$ defined by $(a, b, \rho)$, we use two sample sizes of $n = 100$ and $n = 500$.  For each of the 8 configurations, we conduct a Monte Carlo (MC) simulation with a MC sample size of 1000.

We refer to the procedure proposed in Section 2 as ME-Cor. We compare the proposed procedure to two primary competitors. One competitor is the approach of \cite{ybarra2008small}, which assumes that $Cor(u_{i}, e_{i}) = 0$. We abbreviate the \cite{ybarra2008small} procedure as ``YL''.  We implement estimation and prediction for the model of \cite{ybarra2008small} using the {\tt R} package {\tt saeME}.  The other competitor is the standard estimator and predictor for the traditional \cite{fay1979estimates} model, abbreviated as ``FH.'' This competitor is of practical interest because naive application of the Fay-Herriot model when the covariate is from a sample survey is tempting for its simplicity.  We implement estimation, prediction, and MSPE estimation for the standard Fay-Herriot model using the {\tt R} package {\tt SAE}.  

We do not include the predictor outlined in \cite{ybarra2003} in the simulations for two main reasons. One is that the predictor of \cite{ybarra2003} is not fully developed for the case of a correlation between the measurement error and the sampling error. The other is that we do not view the procedure of \cite{ybarra2003} as a competitor to our approach. Instead, our objective is to build on the predictor of \cite{ybarra2003} and study its properties in more detail.  

We also do not compare our predictor to a predictor for a bivariate model in which the covariate is included as a second response variable. As discussed in Section 1, \cite{franco2019using} considers a bivariate modeling approach. Their approach is equivalent to a structural measurement error model with no covariates.  We prefer to remain in the framework of functional measurement error. Therefore, we do not include a comparison to a predictor based on a bivariate model.

\subsection{  Normal Distributions, Unequal $\Psi_i$ }

We first simulate data with normally distributed random components, as specified in model (\ref{mecor}). We use the unequal $\bm{\Psi}_{i}$ defined in (\ref{unequalpsi}). We compare the alternative estimators and predictors for the case of normal distributions and unequal $\bm{\Psi}_{i}$ in Tables ~\ref{tab1} and ~\ref{tab2}.  
Table~\ref{tab1} contains the MC means and standard deviations of the alternative estimators of the fixed model parameters. The properties of the estimators are similar across the 8 configurations. At the sample size of $n = 500$, the proposed estimator (ME-Cor) is approximately unbiased for the intercept and slope. For $n= 500$, the MC bias of the proposed estimator of $\sigma^{2}_{b}$ is below 0.01 for each combination of $(a,b,\rho)$. 

The proposed estimator of $\sigma^{2}_{b}$ usually has a negative bias when $n = 100$. A negative small-sample bias for the estimator of $\sigma^{2}_{b}$ is not surprising because the estimator does not incorporate a correction for the loss of degrees of freedom associated with estimating $\beta_{0}$ and $\beta_{1}$. An exception to the negative bias occurs when the measurement error exceeds the sampling error and when the correlation between $u_{i}$ and $e_{i}$ is only 0.2. For this configuration, the estimator of $\sigma^{2}_{b}$ has a positive bias at $n = 100$. Increasing the sample size to $n = 500$ rectifies the bias of the estimator of $\sigma^{2}_{b}$ for the configuration with $(a,b,\rho) = (0.75, 0.25, 0.2)$. When $(a, b, \rho) = (0.75, 0.25, 0.2)$, the distribution of the estimator of $\sigma^{2}_{b}$ for $n = 100$ is highly skewed  right and has extreme values. When the sample size increases to $n = 500$, the distribution of the estimator of $\sigma^{2}_{b}$ for this configuration is more unimodal and symmetric. 

The presence of a nontrivial correlation between $u_{i}$ and $e_{i}$ causes the YL estimator to have a negative bias for the intercept and a positive bias for the slope. The YL estimator of $\sigma^{2}_{b}$ has a severe negative bias in the presence of a nonzero correlation between $u_{i}$ and  $e_{i}$. The {\tt R} function {\tt FHme}, used to implement the YL procedure, applies a lower bound of zero to the estimator of $\sigma^{2}_{b}$, and it is apparent from the results that many of the estimates reach the lower bound of zero. 

The measurement error attenuates the FH estimator of the slope toward zero and leads to a positive bias in the estimator of the intercept. The FH estimator of $\sigma^{2}_{b}$ usually has a positive bias because the FH estimator of $\sigma^{2}_{b}$ includes part of the measurement error. For configurations with $a <b$ and $\rho = 0.8$, the covariance structure causes the FH estimator of $\sigma^{2}_{b}$ to have a negative bias.

Table~\ref{tab2} summarizes the empirical properties of the alternative predictors and MSPE estimators. The columns under the heading ``MC MSPE of Predictor'' contain the average MC MSPE's of the alternative predictors, where the average is across areas. The columns under the heading ``MC Mean Est. MSPE'' contain the average MC means of the MSPE estimators for the ME-Cor and FH procedures. The column labeled ``Direct'' indicates the average MC MSPE of the direct estimator, $Y_{i}$. 

The YL predictor is superior to the FH predictor but inferior to the ME-Cor predictor. For all but the configuration with $(a,b,\rho) = (0.25, 0.75, 0.2),$ the YL predictor has MC MSPE exceeding that of the direct estimator. The results for the YL predictor demonstrate the importance of accounting for a correlation between $u_{i}$ and $e_{i}$.  

The properties of the FH predictor depend on the structure of $\bm{\Psi}_{i}$ and are similar for $n = 100$ and $n = 500$. When $\rho = 0.2$ and $a < b$ (measurement error variance is smaller than the sampling variance), the FH predictor is superior to the direct estimator but inferior to the ME-Cor predictor. For all other configurations, the MC MSPE of the FH predictor exceeds the MC MSPE of the direct estimator. This empirical finding echoes a theoretical result in \cite{ybarra2008small} that the Fay-Herriot predictor can have MSPE greater than the variance of the direct estimator if the covariate is measured with error. 

The efficiency of the FH predictor relative to the direct estimator is best when $a < b$ and $\rho = 0.2$. This is reasonable because this configuration most closely approximates a situation where the covariate is measured without error. The MC means of the estimators of $\sigma^{2}_{b}$ in Table~\ref{tab1} provide insight into why the FH predictor is less efficient than the direct estimator for $(a,b,\rho) \neq (0.25, 0.75, 0.2)$. The MC means of the estimators of $\sigma^{2}_{b}$ in Table~\ref{tab1} reveal the impacts of measurement error on the shrinkage parameters for the FH predictor, where the shrinkage parameter is defined as $\hat{\sigma}^{2}_{b}/(\hat{\sigma}^{2}_{b} + \psi_{eei})$. For the scenarios with $(a,b,\rho) = (0.25, 0.75, 0.8)$, the estimator of $\sigma^{2}_{b}$ has a severe negative bias, leading to considerable over-shrinkage toward a covariate that is itself measured with error. For scenarios with $a > b$, the positive bias of the estimator of $\sigma^{2}_{b}$ is overwhelming, and the FH estimator does not exhibit enough shrinkage toward the estimated regression line.

\begin{table}[ht]
	\centering
	\caption{ MC means and standard deviations of estimators of fixed model parameters. True values are $(\beta_{0}, \beta_{1}, \sigma^{2}_{b}) = (1, 2, 0.36)$. Normal distributions; unequal $\bm{\Psi}_{i}$. } \label{tab1} 
	\setlength{\tabcolsep}{5pt} 
	\renewcommand{\arraystretch}{1.5}
	\begin{tabular}{ccccccccccc}
		\hline
		&     & & & &\multicolumn{2}{c}{\underline{ME-Cor}} &\multicolumn{2}{c}{\underline{YL}} & \multicolumn{2}{c}{\underline{FH}} \\
		& $a$ & $b$ &$\rho$& $n$ & MC Mean & MC SD & MC Mean & MC SD & MC Mean & MC SD \\  
		\hline
$\beta_{0}$&   &   &   & 100 & 0.987 & 0.298 & 0.942 & 0.263 & 1.266 & 0.260 \\[-0.1 cm] 
$\beta_{1}$ & 0.250 & 0.750 & 0.200 & 100 & 2.002 & 0.054 & 2.011 & 0.046 & 1.944 & 0.045 \\[-0.1 cm] 
$\sigma^{2}_{b}$ &   &  &   & 100 & 0.351 & 0.236 & 0.083 & 0.172 & 1.050 & 0.291 \\[-0.1 cm] 
 \hline 
$\beta_{0}$&   &   &   & 500 & 0.993 & 0.122 & 0.953 & 0.112 & 1.218 & 0.109 \\[-0.1 cm] 
$\beta_{1}$ & 0.250 & 0.750 & 0.200 & 500 & 2.001 & 0.021 & 2.009 & 0.019 & 1.956 & 0.018 \\[-0.1 cm] 
$\sigma^{2}_{b}$ &   &   &   & 500 & 0.357 & 0.108 & 0.019 & 0.053 & 1.045 & 0.125 \\[-0.1 cm] 
 \hline 
$\beta_{0}$ &   &   &   & 100 & 0.996 & 0.176 & 0.840 & 0.170 & 1.071 & 0.164 \\[-0.1 cm] 
$\beta_{1}$ & 0.250 & 0.750 & 0.800 & 100 & 2.001 & 0.027 & 2.028 & 0.026 & 1.988 & 0.025 \\[-0.1 cm] 
$\sigma^{2}_{b}$ &   &   &  & 100 & 0.344 & 0.106 & 0.000 & 0.000 & 0.058 & 0.073 \\[-0.1 cm] 
 \hline 
$\beta_{0}$ &   &   &   & 500 & 1.001 & 0.077 & 0.810 & 0.079 & 1.084 & 0.075 \\[-0.1 cm] 
$\beta_{1}$ & 0.250 & 0.750 & 0.800 & 500 & 2.000 & 0.013 & 2.038 & 0.013 & 1.983 & 0.012 \\[-0.1 cm] 
$\sigma^{2}_{b}$ &   &   &   & 500 & 0.359 & 0.047 & 0.000 & 0.000 & 0.044 & 0.037 \\[-0.1 cm] 
 \hline 
$\beta_{0}$&   &   &   & 100 & 0.968 & 0.470 & 0.912 & 0.431 & 1.972 & 0.363 \\[-0.1 cm] 
$\beta_{1}$ & 0.750 & 0.250 & 0.200 & 100 & 2.008 & 0.085 & 2.018 & 0.075 & 1.803 & 0.062 \\[-0.1 cm] 
$\sigma^{2}_{b}$ &   &  &   & 100 & 0.377 & 0.375 & 0.175 & 0.335 & 3.410 & 0.568 \\[-0.1 cm] 
 \hline 
$\beta_{0}$ &   &   &   & 500 & 1.002 & 0.186 & 0.957 & 0.163 & 1.833 & 0.151 \\[-0.1 cm] 
$\beta_{1}$ & 0.750 & 0.250 & 0.200 & 500 & 2.000 & 0.031 & 2.009 & 0.027 & 1.838 & 0.024 \\[-0.1 cm] 
$\sigma^{2}_{b}$ &   &  &   & 500 & 0.361 & 0.191 & 0.067 & 0.135 & 3.502 & 0.257 \\[-0.1 cm] 
 \hline 
$\beta_{0}$ &   &   &   & 100 & 0.991 & 0.352 & 0.778 & 0.325 & 1.758 & 0.272 \\[-0.1 cm] 
$\beta_{1}$ & 0.750 & 0.250 & 0.800 & 100 & 2.004 & 0.064 & 2.048 & 0.060 & 1.844 & 0.047 \\[-0.1 cm] 
$\sigma^{2}_{b}$ &   &   &   & 100 & 0.350 & 0.282 & 0.001 & 0.018 & 2.174 & 0.392 \\[-0.1 cm] 
 \hline 
$\beta_{0}$ &   &   &   & 500 & 0.996 & 0.149 & 0.819 & 0.138 & 1.708 & 0.119 \\[-0.1 cm] 
$\beta_{1}$ & 0.750 & 0.250 & 0.800 & 500 & 2.000 & 0.026 & 2.036 & 0.024 & 1.856 & 0.020 \\[-0.1 cm] 
$\sigma^{2}_{b}$ &   &   &   & 500 & 0.360 & 0.135 & 0.000 & 0.000 & 2.205 & 0.169 \\[-0.1 cm] 
		\hline 
	\end{tabular}
\end{table}

\newpage
\clearpage

An important implication of measurement error is that the Fay-Herriot MSPE estimator (FH-MSPE) has a severe negative bias for the MSPE of the Fay-Herriot predictor (FH). When $a < b$ and $\rho = 0.8$, the MC mean of the FH MSPE estimator is more than an order of magnitude lower than the MC MSPE of the FH predictor. A risk of naive application of Fay-Herriot procedures in the presence of measurement error is under-estimation of the MSPE. 

The ME-Cor predictor has smaller MC MSPE than the alternatives considered for all configurations. When $a < b$,  the gain in efficiency from the ME-Cor predictor relative to the direct estimator is greater for $\rho  = 0.2$ than for $\rho = 0.8$. When $a> b$, the opposite pattern holds, as the ratio of the MC MSPE of the direct estimator to the MC MSPE of the ME-Cor predictor is greater for $\rho = 0.8$ than for $\rho = 0.2$. The ME-Cor procedure renders only trivial improvements in efficiency over the direct estimator for configurations with $(a,b,\rho) = (0.25, 0.75, 0.8)$ or $(a,b,\rho) = (0.75, 0.25, 0.2)$.  Increasing $n$ from 100 to 500 has little effect on the properties of the ME-Cor predictor. 

The proposed MSPE estimator (ME-Cor-MSPE) is a good approximation for the MSPE of the ME-Cor predictor (ME-Cor). For each configuration, the average MC mean of the estimated MSPE for the ME-Cor predictor (ME-Cor-MSPE) is close to the average MC MSPE of the ME-Cor predictor (ME-Cor). The simulation results support the predictor and MSPE estimator proposed in Section 2.

\begin{table}[ht]
	\centering
	\caption{ MC MSPE's of alternative predictors (MC MSPE of Predictor) and MC means of estimated MSPE's (MC Mean Est. MSPE) of ME-Cor and FH predictors. Normal distributions; unequal $\bm{\Psi}_{i}$. }\label{tab2} 
	\setlength{\tabcolsep}{5pt} 
	\renewcommand{\arraystretch}{1.5}
	\begin{tabular}{cccccccccc}
		\hline
		& & & & \multicolumn{ 4}{c}{\underline{MC MSPE of Predictor}} & \multicolumn{2}{c}{\underline{MC Mean Est. MSPE}}  \\
		\hline \hline 
		\multicolumn{3}{c}{$(a, b, \rho)$} & $n$ & Direct &   ME-Cor&  YL & FH & ME-Cor-MSPE & FH-MSPE \\[-0.1 cm]  
		\hline
 \multicolumn{3}{c}{(0.250, 0.750, 0.200)} & 100  & 1.010 & 0.748 & 0.759 & 0.809 & 0.744 & 0.499 \\[-0.1 cm]  
  &   &  & 500  & 1.007 & 0.741 & 0.758 & 0.804 & 0.739 & 0.487 \\[-0.1 cm]  
 \hline 
 \multicolumn{3}{c}{(0.250, 0.750, 0.800)} & 100 & 1.010 & 1.002 & 1.105 & 1.584 & 1.001 & 0.087 \\[-0.1 cm]  
   &   &   & 500 & 1.007 & 1.000 & 1.112 & 1.614 & 1.001 & 0.048 \\[-0.1 cm]  
 \hline 
 \multicolumn{3}{c}{(0.750, 0.250, 0.200)} & 100  & 0.337 & 0.335 & 0.345 & 0.357 & 0.334 & 0.302 \\[-0.1 cm]  
   &   &  & 500  & 0.336 & 0.334 & 0.345 & 0.356 & 0.333 & 0.301 \\[-0.1 cm]  
 \hline 
 \multicolumn{3}{c}{(0.750, 0.250, 0.800)} & 100  & 0.334 & 0.214 & 0.440 & 0.559 & 0.215 & 0.285 \\[-0.1 cm]  
   &   &   & 500  & 0.336 & 0.213 & 0.443 & 0.563 & 0.212 & 0.285 \\[-0.1 cm]  
		\hline
	\end{tabular}
\end{table}

\subsection{ Normal Distributions, Equal $\Psi_i = \Psi$}

A special case in which the FH predictor retains reasonable properties occurs in the context of the structural model when the measurement error is uncorrelated with the sampling error and when the measurement error variance is constant \citep{bell2019measurement}.
When the measurement error and sampling error are correlated, the naive Fay-Herriot predictor remains inappropriate, even if the measurement error variance is constant. To illustrate this point, we present simulation results with equal $\bm{\Psi}_{i} = \bm{\Psi}$.

Tables~\ref{tab1equal} and \ref{tab2equal} contain simulation results for $\bm{\Psi}_{i} = \bm{\Psi}_{1}$ for $i = 1,\ldots, n$, where $\bm{\Psi}_{1}$ is defined in (\ref{unequalpsi}). We continue to simulate the errors from normal distributions. The conclusions for equal $\bm{\Psi}_{i}$ are the same as those for unequal $\bm{\Psi}_{i}$. As seen in Table~\ref{tab1equal}, the FH and YL estimators of the fixed parameters remain biased when $\bm{\Psi}_{i} = \bm{\Psi}$. In contrast, the MC means of the proposed estimators of the fixed parameters are close to the true parameter values. In Table~\ref{tab2equal}, the proposed predictor has smaller MC MSPE than the alternatives for all configurations and sample sizes. The Fay-Herriot predictor remains inefficient when $\bm{\Psi}_{i} = \bm{\Psi}$. Also, the FH estimator of the MSPE continues to have a negative bias in the case of equal $\bm{\Psi}_{i}$. The proposed MSPE estimator is nearly unbiased for the MSPE of the ME-Cor predictor.

\subsection{Simulations with $t$ Distributions}

We next simulate data from $t$ distributions. This allows us to assess robustness of the procedure to departures from normality. To simulate data from $t$ distributions, we first generate $\tilde{u}_{i}\stackrel{iid}{\sim} t_{(5)}/\sqrt{5/3}$, $\tilde{e}_{i}\stackrel{iid}{\sim}t_{(5)}/\sqrt{5/3}$, and $\tilde{b}_{i} \stackrel{iid}{\sim} t_{(5)}/\sqrt{5/3}$.  The notation $t_{(5)}/\sqrt{5/3}$ denotes a $t$ distribution with 5 degrees of freedom divided by $\sqrt{5/3}$. The division by $\sqrt{5/3}$ standardizes the variables so that $\tilde{u}_{i}$, $\tilde{e}_{i}$, and $\tilde{b}_{i}$ have zero mean and unit variance. The $\tilde{u}_{i}$, $\tilde{e}_{i}$, and $\tilde{b}_{i}$ are mutually independent.  We then define $(u_{i}, e_{i})' = \bm{\Psi}_{i}^{0.5}(\tilde{u}_{i}, \tilde{e}_{i})'$, where $\bm{\Psi}_{i}^{0.5}$ is the square root matrix of $\bm{\Psi}_{i}$.  We set $b_{i} = \sigma_{b}\tilde{b}_{i}$.  We use the $\bm{\Psi}_{i}$  defined in (\ref{unequalpsi}) with the 4 combinations of $(a,b,\rho)$.

\begin{table}[ht]
	\centering
	\caption{ MC means and standard deviations of estimators of fixed model parameters. True values are $(\beta_{0}, \beta_{1}, \sigma^{2}_{b}) = (1, 2, 0.36)$. Normal distributions, equal $\bm{\Psi}_{i} = \bm{\Psi}$. } \label{tab1equal} 
		\setlength{\tabcolsep}{5pt} 
	\renewcommand{\arraystretch}{1.5}
	\begin{tabular}{ccccccccccc}
		\hline
		&     & & & &\multicolumn{2}{c}{\underline{ME-Cor}} &\multicolumn{2}{c}{\underline{YL}} & \multicolumn{2}{c}{\underline{FH}} \\
		& $a$ & $b$ &$\rho$& $n$ & MC Mean & MC SD & MC Mean & MC SD & MC Mean & MC SD \\ 
		\hline
 $\beta_{0}$ &   &   &   & 100& 0.994 & 0.252 & 0.949 & 0.253 & 1.202 & 0.243 \\[-0.1 cm] 
 $\beta_{1}$ & 0.250 & 0.750 & 0.200 & 100& 2.001 & 0.044 & 2.010 & 0.044 & 1.959 & 0.042 \\[-0.1 cm] 
$\sigma^{2}_{b}$ &   &   &   & 100& 0.336 & 0.228 & 0.090 & 0.143 & 0.999 & 0.248 \\[-0.1 cm] 
  \hline 
$\beta_{0}$ &   &   &   & 500& 0.997 & 0.111 & 0.953 & 0.111 & 1.204 & 0.107 \\[-0.1 cm] 
$\beta_{1}$ & 0.250 & 0.750 & 0.200 & 500& 2.001 & 0.019 & 2.010 & 0.019 & 1.960 & 0.018 \\[-0.1 cm] 
$\sigma^{2}_{b}$ &   &   &   & 500& 0.351 & 0.112 & 0.047 & 0.067 & 0.996 & 0.112 \\[-0.1 cm] 
  \hline 
$\beta_{0}$ &   &   &   & 100& 0.996 & 0.154 & 0.826 & 0.156 & 1.070 & 0.151 \\[-0.1 cm] 
$\beta_{1}$ & 0.250 & 0.750 & 0.800 & 100& 2.001 & 0.027 & 2.035 & 0.027 & 1.986 & 0.026 \\[-0.1 cm] 
$\sigma^{2}_{b}$&   &   &   & 100& 0.341 & 0.101 & 0.000 & 0.000 & 0.027 & 0.052 \\[-0.1 cm] 
  \hline 
$\beta_{0}$ &   &   &   & 500& 0.999 & 0.077 & 0.798 & 0.078 & 1.086 & 0.075 \\[-0.1 cm] 
$\beta_{1}$ & 0.250 & 0.750 & 0.800 & 500& 2.000 & 0.014 & 2.041 & 0.014 & 1.982 & 0.013 \\[-0.1 cm] 
$\sigma^{2}_{b}$ &   &   &   & 500& 0.358 & 0.047 & 0.000 & 0.000 & 0.008 & 0.019 \\[-0.1 cm] 
  \hline 
$\beta_{0}$ &   &   &   & 100& 0.962 & 0.377 & 0.914 & 0.380 & 1.697 & 0.317 \\[-0.1 cm] 
$\beta_{1}$ & 0.750 & 0.250 & 0.200 & 100& 2.005 & 0.065 & 2.015 & 0.066 & 1.863 & 0.053 \\[-0.1 cm] 
$\sigma^{2}_{b}$ &   &   &   & 100& 0.352 & 0.357 & 0.153 & 0.256 & 2.812 & 0.433 \\[-0.1 cm] 
  \hline 
$\beta_{0}$ &   &   &  & 500& 0.988 & 0.148 & 0.948 & 0.149 & 1.604 & 0.129 \\[-0.1 cm] 
$\beta_{1}$ & 0.750 & 0.250 & 0.200 & 500& 2.003 & 0.025 & 2.011 & 0.026 & 1.879 & 0.021 \\[-0.1 cm] 
$\sigma^{2}_{b}$ &   &   &   & 500& 0.342 & 0.196 & 0.078 & 0.120 & 2.839 & 0.192 \\[-0.1 cm] 
  \hline 
$\beta_{0}$ &   &   &   & 100& 0.983 & 0.286 & 0.825 & 0.294 & 1.475 & 0.251 \\[-0.1 cm] 
$\beta_{1}$ & 0.750 & 0.250 & 0.800 & 100& 2.004 & 0.047 & 2.035 & 0.049 & 1.904 & 0.040 \\[-0.1 cm] 
$\sigma^{2}_{b}$ &   &   &   & 100& 0.328 & 0.268 & 0.000 & 0.001 & 1.860 & 0.290 \\[-0.1 cm] 
  \hline 
$\beta_{0}$ &   &   &   & 500& 0.987 & 0.127 & 0.822 & 0.131 & 1.502 & 0.110 \\[-0.1 cm] 
$\beta_{1}$ & 0.750 & 0.250 & 0.800 & 500& 2.002 & 0.023 & 2.036 & 0.023 & 1.896 & 0.019 \\[-0.1 cm]
$\sigma^{2}_{b}$ &   &   &   & 500& 0.352 & 0.142 & 0.000 & 0.000 & 1.857 & 0.136 \\[-0.1 cm] 
		\hline 
	\end{tabular}
\end{table}

\newpage
\clearpage

\begin{table}[ht]
	\centering
	\caption{ MC MSPE's of alternative predictors (MC MSPE of Predictor), and MC means of estimated MSPE's of ME-Cor and FH predictors (MC Mean Est. MSPE). Normal distributions, equal $\bm{\Psi}_{i} = \bm{\Psi}$. }\label{tab2equal} 
		\setlength{\tabcolsep}{5pt} 
	\renewcommand{\arraystretch}{1.5}
	\begin{tabular}{cccccccccc}
		\hline
		& & & & \multicolumn{ 4}{c}{\underline{MC MSPE of Predictor}} & \multicolumn{2}{c}{\underline{MC Mean Est. MSPE}}  \\
		\hline \hline 
			\multicolumn{3}{c}{$(a, b, \rho)$}  & $n$ & Direct &   ME-Cor&  YL & FH & ME-Cor-MSPE & FH-MSPE \\ 
		\hline
   \multicolumn{3}{c}{(0.250, 0.750, 0.200)} & 100 & 0.745 & 0.564 & 0.575 & 0.580 & 0.564 & 0.441 \\  [-0.1 cm] 
    &   &   & 500 & 0.750 & 0.563 & 0.578 & 0.576 & 0.561 & 0.430 \\[-0.1 cm]  
   \hline 
   \multicolumn{3}{c}{(0.250, 0.750, 0.800)} & 100  & 0.747 & 0.744 & 0.842 & 1.291 & 0.746 & 0.067 \\[-0.1 cm]  
     &   &   & 500  & 0.752 & 0.747 & 0.847 & 1.329 & 0.745 & 0.017 \\[-0.1 cm]  
   \hline 
   \multicolumn{3}{c}{(0.750, 0.250, 0.200)} & 100  & 0.249 & 0.247 & 0.255 & 0.255 & 0.248 & 0.230 \\[-0.1 cm]  
     &   &   & 500  & 0.250 & 0.248 & 0.257 & 0.256 & 0.248 & 0.230 \\[-0.1 cm]  
   \hline 
   \multicolumn{3}{c}{(0.750, 0.250, 0.800)} & 100  & 0.248 & 0.164 & 0.327 & 0.372 & 0.163 & 0.222 \\[-0.1 cm]  
     &   &   & 500  & 0.250 & 0.162 & 0.330 & 0.376 & 0.162 & 0.221 \\[-0.1 cm]  
		\hline
	\end{tabular}
\end{table}

The results for the $t$ distribution are presented in Tables~\ref{tab1t} and \ref{tab2t}. Table~\ref{tab1t} contains the MC means and standard deviations of the estimators of the fixed parameters when the random terms are generated from $t$ distributions. Table~\ref{tab2t} contains the MC MSPE's of the predictors as well as the MC means of the estimated MSPE's for the proposed and Fay-Herriot predictors. The results for the $t$ distribution are largely similar to the results for the normal distribution. 

For configurations with $a < b$, the estimator of $\sigma^{2}_{b}$ has a negative bias. The negative bias is expected because the objective function (\ref{ladj}) does not account for the loss of degrees of freedom from estimating regression coefficients. When $a > b$, the heavy tails of the $t$ distribution cause the estimator of $\sigma^{2}_{b}$ to have a positive bias. The positive bias is notable when $n = 100$. Because the random effects have $t$ distributions, the likelihood used to define the estimator of $\sigma^{2}_{b}$ is misspecified for this configuration. Nonetheless, increasing the sample size to $n = 500$ markedly reduces the bias. 

\newpage
\clearpage

\begin{table}[ht]
	\centering
	\caption{ MC means and standard deviations of estimators of fixed model parameters. True values are $(\beta_{0}, \beta_{1}, \sigma^{2}_{b}) = (1, 2, 0.36)$. $t$ distributions; unequal $\bm{\Psi}_{i}$. } \label{tab1t} 
\setlength{\tabcolsep}{5pt} 
	\renewcommand{\arraystretch}{1.5}
	\begin{tabular}{ccccccccccc}
		\hline
		&     & & & &\multicolumn{2}{c}{\underline{ME-Cor}} &\multicolumn{2}{c}{\underline{YL}} & \multicolumn{2}{c}{\underline{FH}} \\
		& $a$ & $b$ &$\rho$& $n$ & MC Mean & MC SD & MC Mean & MC SD & MC Mean & MC SD \\ 
		\hline
$\beta_{0}$ &   &   &  & 100& 0.990 & 0.298 & 0.954 & 0.272 & 1.204 & 0.268 \\[-0.1 cm] 
$\beta_{1}$ & 0.250 & 0.750 & 0.200 & 100& 2.002 & 0.046 & 2.009 & 0.042 & 1.963 & 0.041 \\[-0.1 cm] 
$\sigma^{2}_{b}$&   &   &   & 100& 0.339 & 0.300 & 0.118 & 0.273 & 1.039 & 0.380 \\[-0.1 cm] 
 \hline 
$\beta_{0}$ &   &   &   & 500 & 1.004 & 0.134 & 0.959 & 0.123 & 1.245 & 0.120 \\[-0.1 cm] 
$\beta_{1}$ & 0.250 & 0.750 & 0.200 & 500 & 2.000 & 0.024 & 2.008 & 0.021 & 1.951 & 0.021 \\[-0.1 cm] 
$\sigma^{2}_{b}$ &  &   &   & 500 & 0.354 & 0.149 & 0.037 & 0.093 & 1.037 & 0.169 \\[-0.1 cm] 
 \hline 
$\beta_{0}$ &   &   &   & 100& 1.000 & 0.185 & 0.806 & 0.182 & 1.084 & 0.174 \\[-0.1 cm] 
$\beta_{1}$& 0.250 & 0.750 & 0.800 & 100& 2.000 & 0.030 & 2.039 & 0.030 & 1.984 & 0.028 \\[-0.1 cm] 
$\sigma^{2}_{b}$&   &   &   & 100& 0.349 & 0.154 & 0.000 & 0.000 & 0.070 & 0.114 \\[-0.1 cm] 
 \hline 
$\beta_{0}$ &   &   &   & 500 & 1.001 & 0.080 & 0.817 & 0.080 & 1.082 & 0.077 \\[-0.1 cm] 
$\beta_{1}$ & 0.250 & 0.750 & 0.800 & 500 & 2.000 & 0.014 & 2.037 & 0.014 & 1.984 & 0.013 \\[-0.1 cm]  
$\sigma^{2}_{b}$ &   &   &   & 500 & 0.358 & 0.072 & 0.000 & 0.000 & 0.047 & 0.056 \\[-0.1 cm] 
 \hline 
$\beta_{0}$ &   &   &   & 100& 0.945 & 0.504 & 0.903 & 0.446 & 1.928 & 0.390 \\[-0.1 cm] 
$\beta_{1}$ & 0.750 & 0.250 & 0.200 & 100& 2.011 & 0.097 & 2.020 & 0.086 & 1.807 & 0.073 \\[-0.1 cm] 
$\sigma^{2}_{b}$ &   &   &   & 100& 0.419 & 0.576 & 0.300 & 0.652 & 3.392 & 0.855 \\[-0.1 cm] 
 \hline 
$\beta_{0}$ &   &   &   & 500 & 0.992 & 0.237 & 0.943 & 0.205 & 1.950 & 0.187 \\[-0.1 cm] 
$\beta_{1}$ & 0.750 & 0.250 & 0.200 & 500 & 2.003 & 0.043 & 2.012 & 0.037 & 1.810 & 0.033 \\[-0.1 cm] 
$\sigma^{2}_{b}$ &   &   &   & 500 & 0.369 & 0.306 & 0.131 & 0.267 & 3.431 & 0.387 \\[-0.1 cm] 
 \hline 
$\beta_{0}$ &   &   &   & 100& 0.975 & 0.404 & 0.797 & 0.343 & 1.750 & 0.311 \\[-0.1 cm] 
$\beta_{1}$ & 0.750 & 0.250 & 0.800 & 100& 2.005 & 0.081 & 2.044 & 0.067 & 1.837 & 0.060 \\[-0.1 cm] 
$\sigma^{2}_{b}$ &   &   &  & 100& 0.399 & 0.482 & 0.026 & 0.240 & 2.179 & 0.593 \\[-0.1 cm] 
 \hline 
$\beta_{0}$ &   &   &   & 500 & 0.992 & 0.175 & 0.799 & 0.156 & 1.701 & 0.140 \\[-0.1 cm] 
$\beta_{1}$ & 0.750 & 0.250 & 0.800 & 500 & 2.001 & 0.032 & 2.040 & 0.028 & 1.859 & 0.025 \\[-0.1 cm] 
$\sigma^{2}_{b}$ &   &   &   & 500 & 0.363 & 0.224 & 0.002 & 0.049 & 2.210 & 0.275 \\[-0.1 cm] 
		\hline 
	\end{tabular}
\end{table}

\newpage
\clearpage

The positive bias of the estimator of $\sigma^{2}_{b}$ has minimal impacts on the properties of the predictor. As seen in Table~\ref{tab2t}, the proposed predictor has smaller MC MSPE than the alternatives. The proposed MSPE estimator is a good approximation for the MSPE of the predictor, even when the data are generated from $t$ distributions.  The results for the $t$ distribution support the statement in remark 9 that the predictor and MSPE estimator are robust to the assumption of normality. 

The FH and YL procedures remain inefficient when the data are generated from $t$ distributions. The bias in the estimators of fixed parameters from the FH and YL procedures is much more severe than the bias from the proposed procedure, even when the random terms are generated from $t$ distributions. The bias of the FH and YL estimators propagates  into the predictor, resulting in high prediction MSPE's for the FH and YL procedures. 

\begin{table}[ht]
	\centering
	\caption{ MC MSPE's of alternative predictors (MC MSPE of Predictor), and MC means of estimated MSPE's of ME-Cor and FH predictors (MC Mean Est. MSPE). $t$ distributions; unequal $\bm{\Psi}_{i}$.  }\label{tab2t} 
	\setlength{\tabcolsep}{5pt} 
	\renewcommand{\arraystretch}{1.5}
	\begin{tabular}{cccccccccc}
		\hline
		& & & & \multicolumn{ 4}{c}{\underline{MC MSPE of Predictor}} & \multicolumn{2}{c}{\underline{MC Mean Est. MSPE}}  \\
		\hline \hline 
		\multicolumn{3}{c}{$(a, b, \rho)$} & $n$ & Direct &   ME-Cor&  YL & FH & ME-Cor-MSPE & FH-MSPE \\[-0.1 cm]  
		\hline
  \multicolumn{3}{c}{(0.250, 0.750, 0.200)} & 100  & 0.995 & 0.741 & 0.750 & 0.814 & 0.744 & 0.493 \\[-0.1 cm]  
    &   &   & 500  & 1.004 & 0.738 & 0.754 & 0.802 & 0.739 & 0.484 \\[-0.1 cm]  
  \hline 
 \multicolumn{3}{c}{(0.250, 0.750, 0.800)} & 100  & 0.989 & 0.983 & 1.096 & 1.559 & 1.001 & 0.094 \\[-0.1 cm]  
    &   &  & 500  & 1.011 & 1.004 & 1.115 & 1.614 & 1.001 & 0.049 \\[-0.1 cm]  
  \hline 
 \multicolumn{3}{c}{(0.750, 0.250, 0.200)} & 100  & 0.331 & 0.329 & 0.339 & 0.352 & 0.334 & 0.301 \\[-0.1 cm]  
    &   &   & 500  & 0.334 & 0.332 & 0.342 & 0.353 & 0.333 & 0.301 \\[-0.1 cm]  
  \hline 
  \multicolumn{3}{c}{(0.750, 0.250, 0.800)} & 100  & 0.338 & 0.219 & 0.444 & 0.562 & 0.218 & 0.284 \\[-0.1 cm]  
    &   &  & 500  & 0.337 & 0.215 & 0.445 & 0.565 & 0.213 & 0.284 \\[-0.1 cm]  
		\hline
	\end{tabular}
\end{table}

\subsection{Extended Simulations}

We present extended simulation results in the supplementary material. First, we use a $t_{(3)}$ distribution instead of a $t_{(5)}$ distribution for the random terms. We consider the $t_{(3)}$ distribution because, unlike the $t_{(5)}$ distribution, the $t_{(3)}$ distribution does not have a finite fourth moment. We then consider a distribution where the random terms are distributed as centered and scaled $\chi^{2}_{(3)}$ random variables. We use the $\chi^{2}_{(3)}$ distribution because it is skewed. 

These extended configurations allow us to assess the impacts of skewness and absence of fourth moments on the properties of the proposed procedure. The positive bias of the estimator of $\sigma^{2}_{b}$ is more severe for these configurations because the profile likelihood (\ref{ladj}) is misspecified. The bias for the estimator of $\sigma^{2}_{b}$ has little impact on the other model parameters. The estimators of regression coefficients remain approximately unbiased under the $t_{(3)}$ and $\chi^{2}_{(3)}$ distributions. The predictors remain more efficient than the alternatives considered. Despite the bias for $\sigma^{2}_{b}$, the proposed MSPE estimator continues to provide a reasonable approximation to the MSPE of the predictor. We refer the reader to the supplementary material for further detail. 

We also validate the proposed procedure for a multivariate covariate in the supplementary material. We use two covariates, both of which are measured with error. The estimators of the fixed parameters remain nearly unbiased in the presence of a bivariate covariate. The proposed MSPE estimator is also nearly unbiased for the MSPE of the predictor.

\section{CEAP Data Analysis}

We apply the method proposed in Section 2 to predict mean log sheet and rill erosion in Iowa counties using CEAP data. Iowa has $n=99$ counties as small areas. In CEAP, sheet and rill erosion is measured using a computer model called RUSLE2. A variable that impacts the amount of sheet and rill erosion in a county is the quantity of water runoff. The mean runoff is unknown for the full population of cropland in Iowa. We use the sample mean of runoff obtained from the CEAP survey as the covariate for the small area model. The response is the log of the sample mean of RUSLE2. 

\begin{figure}[ht]
	\centering 
	\includegraphics[scale = 0.35]{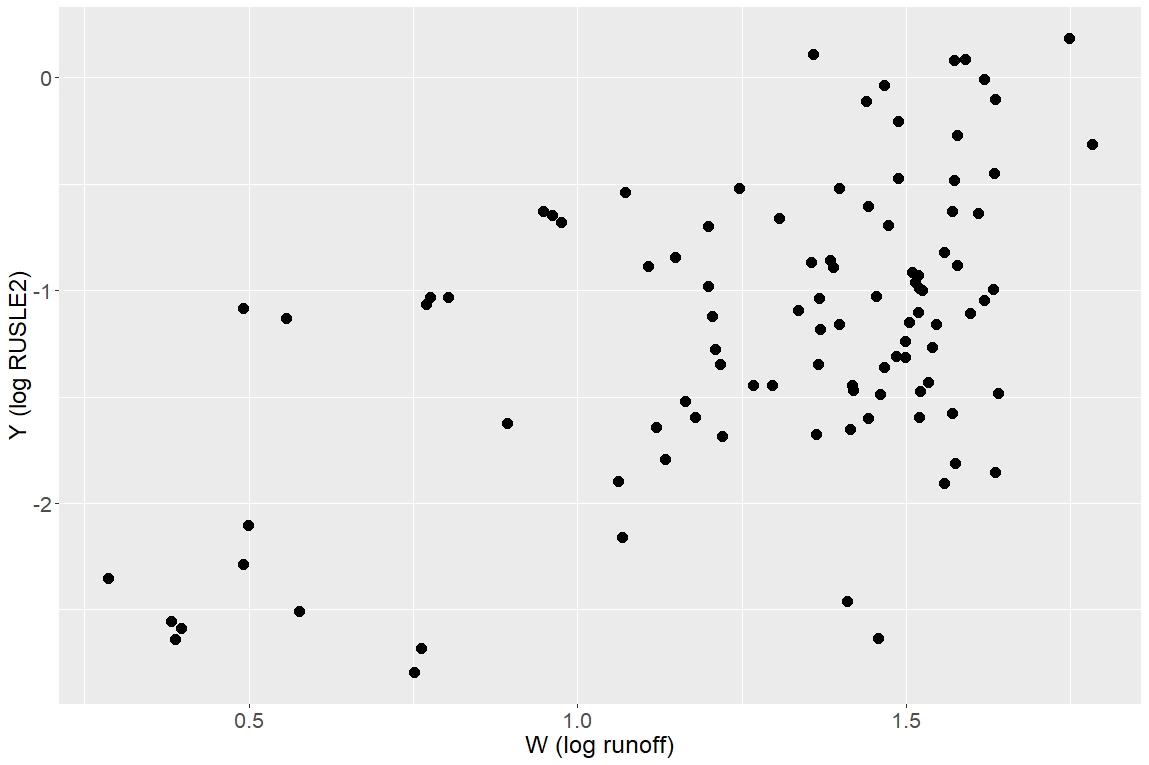}
	\caption{ Scatterplot of $Y_{i} = \mbox{log}(\tilde{Y}_{i})$ on the y-axis against $W_{i} = \mbox{log}(\tilde{W}_{i})$ on the x-axis. } 
	\label{scatterWY1}
\end{figure}

We connect the context and notation of Section 2 to the CEAP data analysis. Let $\tilde{Y}_{i}$ denote the direct estimator of mean RUSLE2 erosion in county $i$, where $i = 1,\ldots, n$. Let $\tilde{W}_{i}$ denote the direct estimator of the mean runoff for county $i$. The direct estimators are defined in Appendix C. The unknown population mean runoff for county $i$ is denoted by $\tilde{x}_{i}$. We define the small area model in the log scale. The response variable  for the small area model is defined by $Y_{i} = \mbox{log}(\tilde{Y}_{i})$. The covariate $W_{i}=\mbox{log}(\tilde{W}_{i})$ is a contaminated measurement of the log of the population mean runoff, defined as $x_{i} = \mbox{log}(\tilde{x}_{i})$.  Figure~\ref{scatterWY1} contains a plot of $Y_{i}$ on the vertical axis against $W_{i}$ on the horizontal axis. The figure exhibits high variation in the association between runoff and RUSLE2. The variation may arise from inherent variability between the counties as well as measurement error in the covariate. The model (\ref{mecor}) accounts for both of these sources of variation.  

The model requires an estimate of $\bm{\Psi}_{i}$, the design variance of $(W_{i}, Y_{i})$. We explain how we estimate the design variance of $(W_{i}, Y_{i})$  in Appendix C. The estimates of $Cor(u_{i}, e_{i})$ range from -0.84 to 1, and the average correlation is 0.18 for the CEAP data. The estimate of the measurement error variance is uniformly smaller than the estimate of the variance of the sampling error for the CEAP data. The ratios of the estimates of $V(u_{i})$ to the estimates of $V(e_{i})$ range from 0.0002 to 0.3436, and the average ratio is 0.090. Although the measurement error is smaller than the sampling error, naive application of Fay-Herriot procedures has the potential to under-estimate the MSPE. 

For this analysis, we treat the direct estimator of $\bm{\Psi}_{i}$ as fixed. An alternative is to use a generalized variance function to smooth the estimator of $\bm{\Psi}_{i}$. For the purpose of this analysis, we think that use of the direct estimator is more illuminating because it enables us to demonstrate the impacts of the structure of $\bm{\Psi}_{i}$ on the efficiency of the predictor.

We assume that the model (\ref{mecor}) holds for $(W_{i}, Y_{i})$ for $i = 1,\ldots, n$. The parameter of interest $\theta_{i}$ represents the mean log RUSLE2 erosion for county $i$, where $i = 1,\ldots, 99$.  We use the procedure of Section 2.2 to estimate the parameters of model (\ref{mecor}). We construct predictors of $\theta_{i}$ and MSPE estimators using the method of Section 2.3. 

Table~\ref{tabestceap} contains the estimate of $\bm{\omega} = (\beta_{0}, \beta_{1}, \sigma^{2}_{b})$ for the CEAP data. The standard errors are the square roots of the diagonal elements of the jackknife covariance matrix defined as  $$\hat{V}_{JK}(\hat{\bm{\omega}}) = \sum_{k = 1}^{n} (\hat{\bm{\omega}}^{(k)} - n^{-1}\sum_{k = 1}^{n}\hat{\bm{\omega}}^{(k)}  )  (\hat{\bm{\omega}}^{(k)} - n^{-1}\sum_{k = 1}^{n}\hat{\bm{\omega}}^{(k)}  )'.$$ The magnitude of the estimate of each parameter is more than double the corresponding standard error.

\begin{table}[ht]
	\centering
	\caption{ Estimates of fixed model parameters and associated standard errors for CEAP data.}\label{tabestceap} 
		\setlength{\tabcolsep}{5pt} 
	\renewcommand{\arraystretch}{1.5}
	\begin{tabular}{cccc}
		\hline
		&$\beta_{0}$ & $\beta_{1}$ &  $\sigma^{2}_{b}$ \\ 
		\hline
		$\hat{\bm{\omega}}$  & -2.541 & 1.053 & 0.283 \\ 
		SE	& 0.250 & 0.185 & 0.041 \\ 	
		\hline
	\end{tabular}
\end{table}

For the CEAP data analysis, several of the estimated MSPE's are negative. We therefore apply a lower bound (LB) to the estimated MSPE for the CEAP analysis. We define the MSPE estimator for the CEAP study by $$\hat{MSPE}_{i,LB} = \hat{MSPE}_{i}I[\hat{MSPE}_{i} >0 ] + (M_{1i}(\hat{\bm{\omega}}) + \hat{M}_{2i,JK})I[ \hat{MSPE}_{i} \leq 0],$$ 
 where $\hat{MSPE}_{i}$ is defined in (\ref{msepred}).

\begin{figure}[ht]
	\centering
	\includegraphics[scale = 0.35]{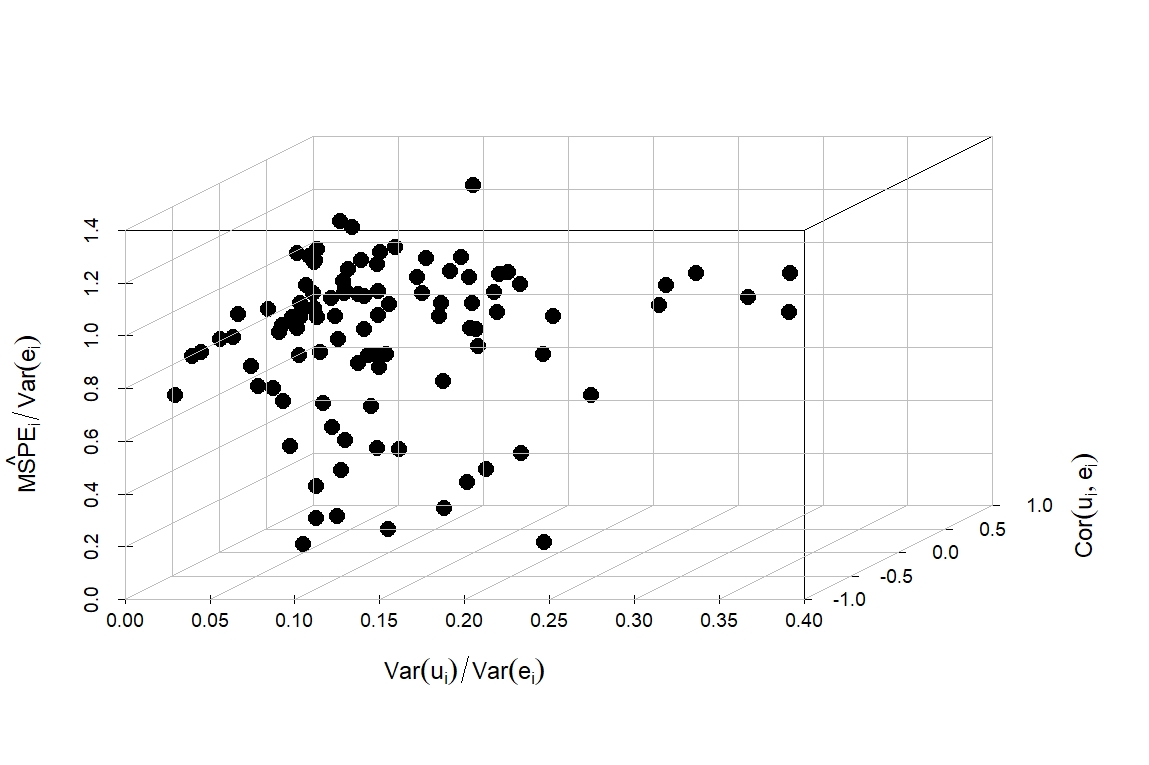} 
	\caption{3-dimensional scatterplot with $Var(u_{i})/Var(e_{i})$ on the x-axis, $Cor(u_{i},e_{i})$ on the y-axis, and the ratio of the estimated MSPE to $Var(e_{i})$ on the z-axis.  }\label{barplotrmse} 
\end{figure}

Figure~\ref{barplotrmse} contains a scatterplot illuminating the relationship between the efficiency gain from prediction and the components of the design covariance matrix. The $x-$axis of the plot contains the ratios of the estimates of $Var(u_{i})$ to the estimates of $Var(e_{i})$. The $y-$axis depicts the estimates of $Cor(u_{i}, e_{i})$. The $z-$axis has the relative mean square prediction error, defined as the ratio of the estimated MSPE of the predictor to the estimated sampling variance of the direct estimator. A ratio below one means that the predictor is more efficient than the direct estimator. 

From Figure~\ref{barplotrmse}, it is apparent that an efficiency gain is attained for most counties. The relative mean square prediction errors range from 0.056 to 1.222, and the average is 0.781. The efficiency gains are often pronounced when $Cor(u_{i}, e_{i})$ and $Var(u_{i})/Var(e_{i})$ are both small. For counties where the estimated MSPE of the predictor exceeds the estimated variance of the direct estimator, $Cor(u_{i}, e_{i})$ tends to be relatively high.  For instance, for the county with a relative mean square error of 1.222, $Cor(u_{i}, e_{i}) = 0.946$. This mirrors the simulation results for the configuration with small measurement error variance and high correlation.  Overall, the estimated efficiency gains from small area modeling are substantial.

\begin{figure}[ht]
\centering
\includegraphics[scale = 0.35]{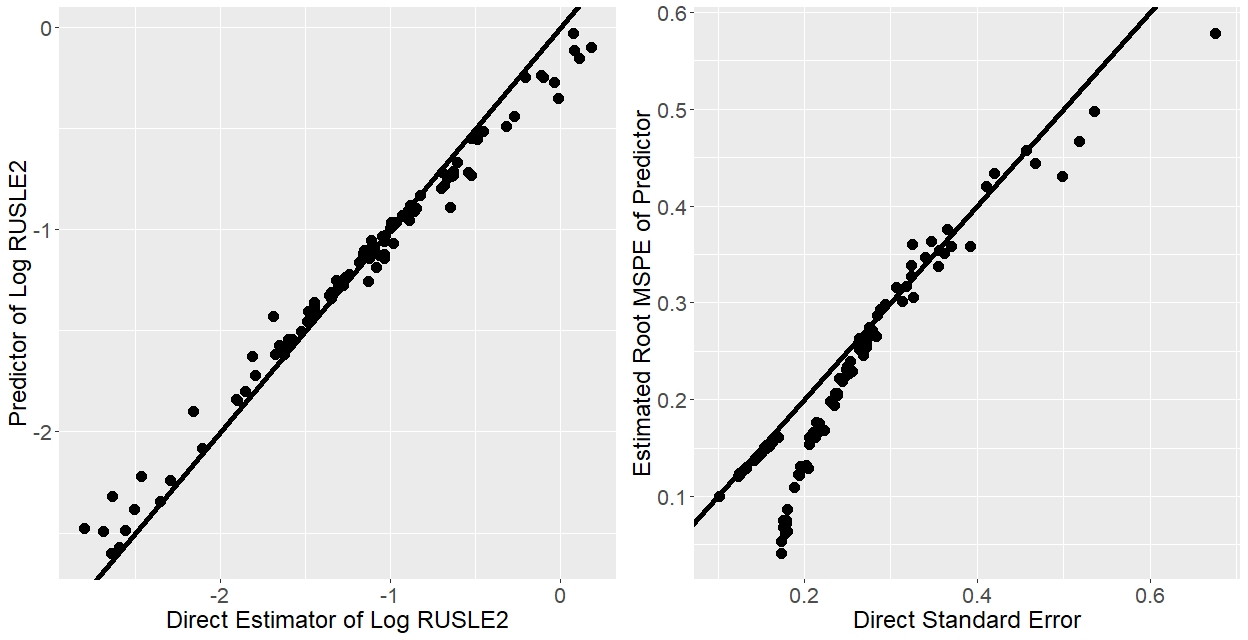}
\caption{(Left) Scatterplot of predictors against direct estimators. (Right) Scatterplot of root MSPE estimators against square roots of estimated variances of direct estimators. }\label{scatterpreddirect} 
\end{figure}

The left panel of Figure~\ref{scatterpreddirect} contains a scatterplot of the predictors on the vertical axis against the direct estimators on the horizontal axis. The line in the plot is the 45-degree line through the origin.  Small area prediction has the expected shrinkage effect. Prediction increases direct estimators that are unusually low and decreases direct estimators that are unusually high.  

The right panel of Figure~\ref{scatterpreddirect} contains a plot of the square roots of the mean square prediction errors against the standard errors of the direct estimators. The line is again the 45-degree line through the origin. For most counties, prediction renders an efficiency gain relative to the direct estimator. The reduction in MSPE from prediction is often substantial.  When the MSPE exceeds the estimated variance of the direct estimator, the estimated loss of efficiency is minimal. 

\section{Discussion}

We conduct an extensive study of the properties of a small area predictor that recognizes a correlation between the measurement error in the covariate and the sampling error in the response. The simulation studies illustrate the dangers of naively applying the Fay-Herriot predictor when the covariate and response are estimators from the same survey.  The Fay-Herriot predictor can be less efficient than the direct estimator, and the corresponding MSPE estimator can have a severe negative bias for the MSPE of the predictor. The problems with the Fay-Herriot predictor persist even when $\bm{\Psi}_{i} = \bm{\Psi}$ for all $i = 1,\ldots, n$.  

The proposed predictor rectifies the problems with the Fay-Herriot predictor and is more efficient than other alternatives considered in the simulations. In both the simulations and the data analysis, the efficiency of the proposed method relative to the direct estimator depends on the nature of $\bm{\Psi}_{i}$. In the CEAP study, runoff varies less within a county than does sheet and rill erosion, and substantial gains in efficiency from small area prediction are observed for most counties.   Without the methodology presented in this paper, the use of runoff as a covariate would be impossible. Our methodology is of general interest beyond the CEAP study. We provide a theoretically sound estimation procedure for use in conjunction with the simple practice of using estimators from related surveys as covariates and response variables. The methodology has potential use in a wide range of applications in the area of official statistics.  

\bibliographystyle{ims}
\bibliography{Bibliography}

\newpage
\clearpage

\section*{Appendix A: Derivation of Conditional Distribution of $e_{i}$ given $v_{i}$}

We provide further detail on the derivation of the distribution of $e_{i}$ given $v_{i}$. Under the assumptions of the model (\ref{mecor}), the bivariate distribution of $(v_{i}, e_{i})$ can be stated as 
\begin{align*}
    \left( \begin{array}{c} b_{i} + e_{i} - \bm{\beta}_{1}'\bm{u}_{i} \\ e_{i}\end{array}\right)  \sim  N_2  \left( \left( \begin{array}{c} 0 \\ 0 \end{array}\right) , \left( \begin{array}{cc} \sigma^{2}_{b} + \sigma^{2}_{\delta i} & \psi_{ee i} - \bm{\beta}_{1}'\bm{\Psi}_{ue i} \\ \psi_{ee i } - \bm{\beta}_{1}'\bm{\Psi}_{uei} & \psi_{ee i} \end{array}\right) \right) ,
\end{align*}
where we have used the definition of $v_{i}$ as $v_{i} = b_{i} + e_{i} - \bm{\beta}_{1}'\bm{u}_{i}$.   By standard properties of the  bivariate normal distribution, 
\begin{align*}
    E(e_{i} \mid v_{i}) = 0 + \frac{ \psi_{ee i } - \bm{\beta}_{1}'\bm{\Psi}_{uei}}{\sigma^{2}_{b} + \sigma^{2}_{\delta i}}v_{i},
\end{align*}
and
\begin{align*}
    Var(e_{i} \mid v_{i}) = \psi_{ee i} - \frac{ (\psi_{ee i} - \bm{\beta}_{1}'\bm{\Psi}_{ue i})^{2}}{\sigma^{2}_{b} + \sigma^{2}_{\delta i}}. 
\end{align*}

\section*{Appendix B: Statistical Properties of Estimators and Predictors}

We state and prove Theorems 1 and 2 for the case of univariate $x_{i}$. For the univariate case, we parametrize $\bm{\Psi}_{i}$ as 
\begin{align*}
\bm{\Psi}_{i} = \left( \begin{array}{cc}  \psi^{2}_{ui} & \psi_{uei} \\ \psi_{uei} & \psi^{2}_{ei} \end{array}\right). 
\end{align*}
We denote the univariate slope and estimator of $x_{i}$ as $\beta_{1}$ and $W_{i}$, respectively. 

\vspace{0.5cm}

{\it Theorem 1:} Assume the following assumptions hold:
\begin{itemize}
	\item[A.1] $\lim_{n\rightarrow\infty} n^{-1}\sum_{i=1}^{n}x_{i} = \mu_{x}$, where $0 < \mu_{x} < \infty$. 
	\item[A.2] $\lim_{n\rightarrow\infty} n^{-1}\sum_{i=1}^{n}x_{i}^{2} = \delta^{2}_{x}$, where $0 < \delta^{2}_{x} < \infty$.
	\item[A.3] $\lim_{n\rightarrow\infty} n^{-1}\sum_{i=1}^{n}(\psi^{2}_{ui}, \psi^{2}_{ei}, \psi_{uei}  ) = (\psi^{2}_{u,\infty}, \psi^{2}_{e,\infty}, \psi_{ue,\infty})$, where $0 < \psi^{2}_{u,\infty}, \psi^{2}_{e,\infty}, \psi_{ue,\infty} < \infty$. 
	\item[A.4] $\lim_{n\rightarrow\infty} n^{-1}\sum_{i=1}^{n}(\psi^{2}_{ui}x_{i}, \psi^{2}_{ui}x_{i}^{2}, \psi^{2}_{ei}x_{i}^{2}, x_{i}\psi_{uei}, x_{i}^{2}\psi_{uei}, x_{i}\psi_{uei}, x_{i}^{2}\psi_{uei}, \psi^{2}_{uei}, \psi^{4}_{ui} )$ = \newline $(c_{1,\infty}, c_{2,\infty}, c_{3,\infty}, c_{4,\infty}, c_{5,\infty}, c_{6,\infty}, c_{7,\infty}, c_{8, \infty}, c_{9, \infty})$, where $ 0 < c_{k,\infty} < \infty$ for $k = 1,\ldots, 9$. 
	\item[A.5] $\delta^{2}_{x} - \mu_{x}^{2} \neq 0$. 
\end{itemize}
Then, $(\hat{\beta}_{0}, \hat{\beta}_{1})\stackrel{p}{\rightarrow} (\beta_{0}, \beta_{1})$. 

\vspace{0.5cm}

{\it Proof:}  One can express $(\hat{\beta}_{0}, \hat{\beta}_{1})$ as $g(\zeta_{1}, \zeta_{2}, \zeta_{3}, \zeta_{4})$, where $\zeta_{k}$ is the univariate version of the vector $\bm{\zeta}_{k}$ defined in Section 2.2. Let $\zeta_{k,\infty}$ denote the probability limit of $\zeta_{k}$ for $k = 1,2,3,4$. By the law of large numbers and assumptions A.1-A.4, $\zeta_{k,\infty}$ is finite. By assumption A.5,  $g$ is a continuous function on a closed set containing $\zeta_{k,\infty}$, $k = 1,2,3,4$. The result then follows from the continuous mapping theorem. We present a more rigorous proof in the supplementary material.  

\vspace{0.5cm}

{\it Theorem 2:}  Let $\bm{\omega} = (\omega_{1}, \omega_{2}, \omega_{3})' = (\beta_{0}, \beta_{1}, \sigma^{2}_{b})'$ and let $\hat{\bm{\omega}} = (\hat{\omega}_{1}, \hat{\omega}_{2}, \hat{\omega}_{3})' = (\hat{\beta}_{0}, \hat{\beta}_{1}, \hat{\sigma}^{2}_{b})'$. Define 
\begin{align*}
S(\omega_{3} \mid \omega_{1}, \omega_{2}) = \frac{1}{2n}\sum_{i = 1}^{n}\frac{v_{i}^{2}}{(\sigma^{2}_{b} + \sigma^{2}_{\delta i})^{2} } - \frac{1}{2 n}\sum_{i=1}^{n}\frac{1}{\sigma^{2}_{b} + \sigma^{2}_{\delta i}},
\end{align*}
where $v_{i} = Y_{i} - \beta_{0}- \beta_{1}W_{i}$. Assume that for any  $(\omega_{1}, \omega_{2}, \omega_{3})$ in the parameter space for $(\beta_{0}, \beta_{1}, \sigma^{2}_{b})$, 
\begin{align*}
\frac{ \partial S(\omega_{3} \mid \omega_{1} , \omega_{2})}{ \partial \omega_{k}  }  = O_{p}(1), \quad k = 1,2,3,
\end{align*}

Then $\hat{\sigma}^{2}_{b}\stackrel{p}{\rightarrow}\sigma^{2}_{b}$. 

\vspace{0.5cm}

{\it Proof: }  See the supplementary material.

\section*{Appendix C: Direct Estimators for CEAP}

We explain how we estimate $\bm{\Psi}_{i}$ for the CEAP data.  We let $\{j = 1,\ldots, n_{i}\}$ denote the index set for the $n_{i}$ elements in the sample for county $i$.  As in \cite{berg2019semiparametric}, we approximate the CEAP sample for a county as a Poisson sample.  The inclusion probability for unit $j$ in county $i$, denoted $\pi_{ij}$, is  defined in \cite{berg2019semiparametric}.  In CEAP operations, replication variance estimation procedures are used. We do not have access to the replicate weights for this analysis. The Poisson approximation preserves the calculated inclusion probabilities and enables a simple variance estimation procedure that is adequate for the purpose of this study.

We let $\tilde{y}_{ij}$ and $\tilde{w}_{ij}$, respectively, denote the values for RUSLE2 and runoff for element $j$ in county $i$. Then, we define $\bar{\tilde{v}}_{i.} = (\tilde{Y}_{i}, \tilde{W}_{i}) = w_{i.}^{-1}\sum_{j  = 1}^{n_{i}}\pi_{ij}^{-1}(\tilde{y}_{ij}, \tilde{w}_{ij})$, where $w_{i.} = \sum_{j = 1}^{n_{i}} \pi_{ij}^{-1}$. We define an estimate of the within-county variance for county $i$ by 
\begin{align*}
\bm{\Sigma}_{i} = \frac{1}{w_{i.}^{2}} \sum_{j = 1}^{n_{i}}\pi_{ij}^{-1}(\pi_{ij}^{-1} - 1)(\tilde{v}_{ij} - \bar{\tilde{v}}_{i.})(\tilde{v}_{ij} - \bar{\tilde{v}}_{i.})',
\end{align*}
where  $\tilde{v}_{ij} = (\tilde{w}_{ij}, \tilde{y}_{ij})'$. By the delta method, an estimator of the sampling variance of $( W_{i}, Y_{i})'$ is then 
\begin{align*}
\bm{\Psi}_{i} = \mbox{diag}(1/\tilde{W}_{i}, 1/\tilde{Y}_{i}) \, \bm{\Sigma}_{i} \, \mbox{diag}(1/\tilde{W}_{i}, 1/\tilde{Y}_{i}). 
\end{align*}

\newpage
\clearpage

\setcounter{secnumdepth}{0}

\begin{center}
Supplement to ``An application of a small area procedure with correlation between measurement error and sampling error to the Conservation Effects Assessment Project''
\end{center}

In this supplement, we prove Theorems 1-3 of the main document and present extended simulation results. The proofs are given in Section 1. The simulation output is provided in Sections 2-3.

\section{Proofs}

\subsection{Proof of Theorem 1}

The proof of Theorem 1 relies heavily on Theorem 1.14 (ii)  of \cite{shao2003mathematical}. For ease of reference for the reader, we state Theorem 1.14 (ii) of \cite{shao2003mathematical} as Lemma 1 below:

\vspace{0.5cm}

{\it Lemma 1 (Theorem 1.14 (ii) of \cite{shao2003mathematical} )}: Let $X_{1}, X_{2},...$ be independent random variables with finite expectations. If there is a constant $p \in [1,2]$ such that 
\begin{align*}
lim_{n\rightarrow\infty} n^{-p}\sum_{i = 1}^{n}E|X_{i}|^{p} = 0,  
\end{align*}
then
\begin{align*}
\frac{1}{n}\sum_{i=1}^{n}(X_{i} - E(X_{i}))\stackrel{p}{\rightarrow} 0. 
\end{align*}

\vspace{0.5cm}

{\it Proof of Theorem 1}: We now prove that $(\hat{\beta}_{0}, \hat{\beta}_{1})\stackrel{p}{\rightarrow} (\beta_{0}, \beta_{1})$. As notation, we let $\zeta_{k}$ be the univariate version of the vector $\bm{\zeta}_{k}$ defined in Section 2.2 of the main document for $k = 1,2,3,4$. Then, $\zeta_{1} = n^{-1}\sum_{i = 1}^{n} Y_{i} W_{i} - n^{-1}\sum_{i = 1}^{n} \psi_{uei}$, $\zeta_{2} = n^{-1}\sum_{i=1}^{n}Y_{i}$, $\zeta_{3} = n^{-1}\sum_{i=1}^{n}W_{i}$, and $\zeta_{4}= n^{-1}\sum_{i=1}^{n}W_{i}^{2} - n^{-1}\sum_{i=1}^{n}\psi^{2}_{ui}$. We first show that $\zeta_{1}\stackrel{p}{\rightarrow} \beta_{0}\mu_{x}+\beta_{1}\delta^{2}_{x}$. Define $\eta_{i} = Y_{i}W_{i}$. Then, 
\begin{align*}
\eta_{i}& = (\beta_{0} + \beta_{1}x_{i} + b_{i} + e_{i})(x_{i} + u_{i}) \\ \nonumber 
		& = \eta_{i,f} + \eta_{i,R1} + \eta_{i,R2},
\end{align*} 
where $\eta_{i,f}= \beta_{0}x_{i} + \beta_{1}x_{i}^{2}$, $\eta_{i,R1} = (\beta_{0}+ \beta_{1}x_{i})u_{i} + e_{i}x_{i} + b_{i}(u_{i} +x_{i})$, and $\eta_{i,R2} = e_{i}u_{i}$. 

By A.1, $lim_{n\rightarrow\infty} n^{-1}\sum_{i = 1}^{n} \eta_{i,f} = \beta_{0}\mu_{x}+ \beta_{1}\delta^{2}_{x}$. The $\{\eta_{i,R2}: i = 1,\ldots, n\}$ are independent and $E(\eta_{i,R2})= \psi_{ue,i}$. By normality, $E|\eta_{i,R2}|^{2} = \psi^{2}_{ui}\psi^{2}_{ei} + 2\psi_{uei}^{2}$. 

\noindent By A.3, $lim_{n\rightarrow\infty} n^{-2}\sum_{i=1}^{n}E|\eta_{i,R2}|^{2} = 0$. Lemma 1 with $p = 2$ implies that $n^{-1}\sum_{i=1}^{n}(\eta_{i,R2} - \psi_{uei}) \stackrel{p}{\rightarrow}0$. The $\{\eta_{i,R1}: i = 1,\ldots, n\}$ are independent, $E(\eta_{i,R1}) = 0$, and $$E|\eta_{i,R1}|^{2} = V(\eta_{i,R1}) = (\beta_{0} + \beta_{1}x_{i})^{2}\psi^{2}_{ui} + x_{i}^{2}\psi^{2}_{ei} + 2x_{i}(\beta_{0} + \beta_{1}x_{i})\psi_{uei} + \sigma^{2}_{b}(\psi^{2}_{ui} + x_{i}^{2}).$$ By A.4 and lemma 1 with $p = 2$, $lim_{n\rightarrow\infty} n^{-1}\sum_{i=1}^{n}\eta_{i,R1}= 0$. The continuous mapping theorem then implies that $n^{-1}\sum_{i=1}^{n}Y_{i}W_{i}\stackrel{p}{\rightarrow} \beta_{0}\mu_{x}+ \beta_{1}\delta^{2}_{x}$. Similarly, let $\eta_{2i} = W_{i}^{2}$. Then, 
\begin{align*}
\eta_{2i}= x_{i}^{2} + 2x_{i}u_{i} + u_{i}^{2}. 
\end{align*}

By A.1, $n^{-1}\sum_{i=1}^{n}x_{i}^{2} = \delta^{2}_{x}$. By A.4 and lemma 1 with $p = 2$, $lim_{n\rightarrow\infty}n^{-1}\sum_{i=1}^{n}x_{i}u_{i} \stackrel{p}{\rightarrow}0$. By normality, $E|u_{i}|^{4} = 3\psi^{4}_{ui}$. Then, A.4 and lemma 1 imply that $lim_{n\rightarrow\infty}n^{-1}\sum_{i = 1}^{n}(u_{i}^{2} - \psi^{2}_{ui}) \stackrel{p}{\rightarrow} 0$. The continuous mapping theorem then implies that 
\begin{align*}
\zeta_{4} = n^{-1}\sum_{i = 1}^{n}\eta_{2i}  - n^{-1}\sum_{i=1}^{n}\psi^{2}_{ui}  \stackrel{p}{\rightarrow} \delta^{2}_{x}. 
\end{align*} 
Similar arguments imply that 
\begin{align*}
\zeta_{2} \stackrel{p}{\rightarrow} \beta_{0}+ \beta_{1}\mu_{x} : = \zeta_{2,\infty},
\end{align*}
and
\begin{align*}
\zeta_{3} \stackrel{p}{\rightarrow} \mu_{x}: = \zeta_{3,\infty}. 
\end{align*}

We have now established that 
\begin{align*}
\zeta_{1} \stackrel{p}{\rightarrow} \beta_{0}\mu_{x}+ \beta_{1}\delta^{2}_{x} : = \zeta_{1,\infty}, \\ \nonumber 
\zeta_{2} \stackrel{p}{\rightarrow} \beta_{0} +\beta_{1}\mu_{x} : = \zeta_{2,\infty}, \\ \nonumber 
\zeta_{3} \stackrel{p}{\rightarrow} \mu_{x}: = \zeta_{3,\infty},
\end{align*}
and
\begin{align*}
\zeta_{4} \stackrel{p}{\rightarrow} \delta^{2}_{x}:=\zeta_{4,\infty}.
\end{align*}
We can express the function defining $(\hat{\beta}_{0}, \hat{\beta}_{1})'$ as 
\begin{align*}
(\hat{\beta}_{0}, \hat{\beta}_{1})' = (g_{1}(\zeta_{1}, \zeta_{2}, \zeta_{3}, \zeta_{4}), g_{2}(\zeta_{1}, \zeta_{2}, \zeta_{3}, \zeta_{4}) )',
\end{align*}
where 
\begin{align*}
g_{1}(\zeta_{1}, \zeta_{2}, \zeta_{3}, \zeta_{4}) = \frac{1}{\zeta_{4} - \zeta_{3}^{2}}(\zeta_{4}\zeta_{2} - \zeta_{3}\zeta_{1}),  
\end{align*}
and
\begin{align*} 
g_{2}(\zeta_{1}, \zeta_{2}, \zeta_{3}, \zeta_{4}) = \frac{1}{\zeta_{4} - \zeta_{3}^{2}}(\zeta_{1} - \zeta_{3}\zeta_{2}).
\end{align*}

Consider the function defined as $g_{1}(\zeta_{1,\infty}, \zeta_{2,\infty}, \zeta_{3,\infty}, \zeta_{4,\infty})$. By A.5, $\zeta_{4,\infty} - \zeta_{3,\infty}^{2}  = \delta^{2}_{x} - \mu_{x}^{2}\neq 0$. Now, 
\begin{align*}
\zeta_{4,\infty}\zeta_{2,\infty} - \zeta_{3,\infty}\zeta_{1,\infty} & = \delta^{2}_{x}(\beta_{0}+ \beta_{1}\mu_{x}) - \mu_{x}(\beta_{0}\mu_{x}+ \beta_{1}\delta^{2}_{x}) \\ \nonumber & = \beta_{0}(\delta^{2}_{x} - \mu_{x}^{2}) , 
\end{align*}
and
\begin{align*}
\zeta_{1,\infty} - \zeta_{3,\infty}\zeta_{2,\infty} = \beta_{1}(\delta_{x}^{2} - \mu_{x}^{2}). 
\end{align*}
This verifies that 
\begin{align*}
g_{1}(\zeta_{1,\infty}, \zeta_{2,\infty}, \zeta_{3,\infty}, \zeta_{4,\infty}) = \beta_{0},
\end{align*}
and
\begin{align*}
g_{2}(\zeta_{1,\infty}, \zeta_{2,\infty}, \zeta_{3,\infty}, \zeta_{4,\infty}) = \beta_{1}. 
\end{align*}

By A.5, $g_{1}$ and $g_{2}$ are continuous in a closed set containing $(\zeta_{1,\infty}, \zeta_{2,\infty}, \zeta_{3,\infty}, \zeta_{4,\infty})$. The continuous mapping theorem then implies that 
\begin{align*}
g_{1}(\zeta_{1}, \zeta_{2}, \zeta_{3}, \zeta_{4})\stackrel{p}{\rightarrow} \beta_{0}, 
\end{align*}
and
\begin{align*}
g_{2}(\zeta_{1}, \zeta_{2}, \zeta_{3}, \zeta_{4})\stackrel{p}{\rightarrow} \beta_{1}. 
\end{align*}
This completes the proof of Theorem 1.

\subsection{Proof of Theorem 2}

Let $\tilde{\sigma}_{b}^{2}$ be the root of the score equation such that  $$0 = S(\tilde{\sigma}^{2}_{b} \mid \beta_{0}, \beta_{1}).$$ Note that  $\tilde{\sigma}^{2}_{b}$ is the maximum likelihood estimator that would be obtained if $\beta_{0}$ and $\beta_{1}$ were known.  By standard maximum likelihood theory, $\tilde{\sigma}^{2}_{b} \stackrel{p}{\rightarrow}  \sigma^{2}_{b}$. We expand the objective function defining $\hat{\sigma}^{2}_{b}$ around $\tilde{\sigma}^{2}_{b}$. A Taylor approximation gives
\begin{align*}
 0 &= S(\hat{\sigma}^{2}_{b} \mid \hat{\beta}_{0}, \hat{\beta}_{1})  \\ \nonumber 
  & =  S(\tilde{\sigma}^{2}_{b} \mid \beta_{0}, \beta_{1}) + \left( \frac{ \partial S(\tilde{\sigma}^{*2}_{b}\mid \beta_{0}^{*}, \beta_{1}^{*} ) }{\partial \bm{\beta}}\right)'(\hat{\bm{\beta}} - \bm{\beta} )  \\& + \frac{ \partial S(\tilde{\sigma}^{*2}_{b} \mid \beta_{0}^{*}, \beta_{1}^{*})}{\partial \sigma^{2}_{b}}(\hat{\sigma}^{2}_{b} - \tilde{\sigma}^{2}_{b}) \\ \nonumber 
  & =\left( \frac{ \partial S(\tilde{\sigma}^{*2}_{b}\mid \beta_{0}^{*}, \beta_{1}^{*}) }{\partial \bm{\beta}}\right)'(\hat{\bm{\beta}} - \bm{\beta} )  \\& + \frac{ \partial S(\tilde{\sigma}^{*2}_{b} \mid \beta_{0}^{*}, \beta_{1}^{*})}{\partial \sigma^{2}_{b}}(\hat{\sigma}^{2}_{b} - \tilde{\sigma}^{2}_{b})  ,  
\end{align*}  
where $\bm{\beta}^{*} = (\beta_{0}^{*}, \beta_{1}^{*})'$ is on the line segment between $\hat{\bm{\beta}} = (\hat{\beta}_{0}, \hat{\beta}_{1})'$ and $\bm{\beta}= (\beta_{0}, \beta_{1})'$, and $\tilde{\sigma}^{*2}_{b}$ is on the line segment between $\hat{\sigma}^{2}_{b}$ and $\tilde{\sigma}^{2}_{b}$. 

We only require a first order approximation  because we have already proved the consistency of $\hat{\bm{\beta}}$ in Theorem 1. We use the notation $\partial S(\tilde{\sigma}^{*2}_{b} \mid \beta_{0}^{*}, \beta_{1}^{*})/(\partial\sigma^{2}_{b})$ to denote the partial derivative of $S(\sigma^{2}_{b} \mid \beta_{0}, \beta_{1})$ with respect to $\sigma^{2}_{b}$, evaluated at $(\beta_{0}^{*}, \beta_{1}^{*}, \tilde{\sigma}^{*2}_{b})$. Similarly,
$\partial S(\tilde{\sigma}^{*2}_{b}\mid \beta_{0}^{*}, \beta_{1}^{*})/(\partial \bm{\beta})$ denotes the partial derivative of $S(\sigma^{2}_{b} \mid \beta_{0}, \beta_{1})$ with respect to $(\beta_{0}, \beta_{1})$ evaluated at $(\beta_{0}^{*}, \beta_{1}^{*}, \sigma^{*2}_{b})$.

Then, by the conditions in the statement of Theorem 2, 
\begin{align}\label{dev} 
(\hat{\sigma}^{2}_{b}  - \tilde{\sigma}^{2}_{b})   =  -\frac{D_{1}}{D_{2}}(\hat{\bm{\beta}} - \bm{\beta}) = o_{p}(1),
\end{align}
where 
\begin{align*}
    D_{1} & = \frac{ \partial S(\tilde{\sigma}^{*2}_{b}\mid \beta_{0}^{*}, \beta_{1}^{*}) }{\partial \bm{\beta}},
\end{align*}
and
\begin{align*}
 D_{2} & = \frac{ \partial S(\tilde{\sigma}^{*2}_{b} \mid \beta_{0}^{*}, \beta_{1}^{*})}{\partial \sigma^{2}_{b}}.
\end{align*}
The  right side of (\ref{dev}) is $o_{p}(1)$ by Theorem 2. Then, $\hat{\sigma}^{2}_{b} - \tilde{\sigma}^{2}_{b} = o_{p}(1)$, and $\hat{\sigma}^{2}_{b} - \sigma^{2}_{b} = \hat{\sigma}^{2}_{b} - \tilde{\sigma}^{2}_{b} + \tilde{\sigma}^{2}_{b}- \sigma^{2}_{b} = o_{p}(1)$. This completes the proof of Theorem 2.

\section{Extended Simulation Results}

\renewcommand{\theequation}{S.\arabic{equation}}
\setcounter{equation}{0}
\renewcommand{\thelemma}{S.\arabic{lemma}}
\renewcommand{\thedefinition}{S.\arabic{definition}}
\setcounter{table}{0}
\renewcommand{\thetable}{S.\arabic{table}}
\setcounter{figure}{0}
\renewcommand{\thefigure}{S.\arabic{figure}}

\subsection{ Random components are multiples of $t_{(3)}$ distributions}

We generate the random effects as $b_{i} = \sigma_{b}r_{i}$, where $r_{i}\stackrel{iid}{\sim}t_{(3)}/\sqrt{3}$, and we generate $(u_{i}, e_{i})'$ as $\bm{\Psi}_{i}^{0.5}(\tilde{u}_{i}, \tilde{e}_{i})$, where $\tilde{u}_{i}$ and $\tilde{e}_{i}$  are independently distributed as $t_{(3)}/\sqrt{3}$ random variables. Tables~\ref{tabt32} and~\ref{tabt31} present the results. For this simulation model, the random terms do not have finite fourth moments. Nonetheless, the properties of the proposed  predictors and MSPE estimators mostly remain stable. The proposed estimators of regression coefficients are nearly unbiased. The heavy tails of the $t_{(3)}$ distribution exacerbate the bias of the estimator of  $\sigma^{2}_{b}$. The bias for the estimator of $\sigma^{2}_{b}$ has minimal impacts on the properties of the predictors and MSPE estimators. The proposed predictors are more efficient than the alternatives. The proposed MSPE estimator is a reasonable approximation for the MSPE of the predictor.

\subsection{Random components are multiples of $\chi^{2}_{(3)}$ distributions} 

We generate the random effects as $b_{i} = \sigma_{b}r_{i}$, where $r_{i}\stackrel{iid}{\sim}(\chi^{2}_{(3)} - 3)/\sqrt{6}$. Similarly, we generate $(u_{i}, e_{i})'$ as $\bm{\Psi}_{i}^{0.5}(\tilde{u}_{i}, \tilde{e}_{i})$, 
where $\tilde{u}_{i}$ and $\tilde{e}_{i}$ have independent $\chi^{2}_{3}$ distributions, centered and scaled to have mean zero and unit variance.  Tables~\ref{tabchi1} and ~\ref{tabchi2} present the results. For this simulation model, the random effects have skewed distributions. Despite the skewness, the estimators of fixed parameters and predictors have reasonable properties. The estimators of regression coefficients are nearly unbiased. The estimator of $\sigma^{2}_{b}$ can have a positive bias when $n = 100$ and $a > b$. Increasing the sample size to 500 reduces the magnitude of the bias substantially.  Despite the bias for the estimator of $\sigma^{2}_{b}$, the proposed predictors are more efficient than the alternatives, and the proposed MSPE estimator is a reasonable approximation for the MSPE of the predictor.

\begin{table}[ht]
\centering 
	\caption{ MC MSPE's of alternative predictors (MC MSPE of Predictor), and MC means of estimated MSPE's of ME-Cor and FH predictors (MC Mean Est. MSPE).  $t_{(3)}$ distributions; unequal $\bm{\Psi}_{i}$.  }\label{tabt32} 
	\setlength{\tabcolsep}{5pt} 
	\renewcommand{\arraystretch}{1.5}
	\begin{tabular}{cccccccccc}
		\hline
	&	 & & & \multicolumn{ 4}{c}{\underline{MC MSPE of Predictor}} & \multicolumn{2}{c}{\underline{MC Mean Est. MSPE}}  \\
		\hline \hline 
	\multicolumn{3}{c}{$(a, b, \rho)$} & $n$ & Direct &   ME-Cor&  YL & FH & ME-Cor-MSPE & FH-MSPE \\[-0.1 cm] 
		\hline
    \multicolumn{3}{c}{(0.250, 0.750, 0.200)} &   100  & 0.939 & 0.705 & 0.707 & 0.804 & 0.749 & 0.447 \\[-0.1 cm]  
      &  &  & 500  & 0.988 & 0.731 & 0.741 & 0.805 & 0.740 & 0.461 \\[-0.1 cm]  
 \hline 
    \multicolumn{3}{c}{(0.250, 0.750, 0.800)} & 100  & 1.015 & 1.012 & 1.116 & 1.543 & 1.001 & 0.096 \\[-0.1 cm]  
      &   &   & 500  & 0.977 & 0.972 & 1.084 & 1.555 & 1.001 & 0.051 \\[-0.1 cm]  
  \hline 
 \multicolumn{3}{c}{(0.750, 0.250, 0.200)} & 100  & 0.320 & 0.318 & 0.325 & 0.340 & 0.334 & 0.297 \\[-0.1 cm]  
      &  &  & 500  & 0.333 & 0.330 & 0.339 & 0.352 & 0.334 & 0.298 \\[-0.1 cm] 
  \hline 
 \multicolumn{3}{c}{(0.750, 0.250, 0.800)} & 100  & 0.334 & 0.222 & 0.431 & 0.569 & 0.223 & 0.279 \\[-0.1 cm]  
     &   &   & 500  & 0.330 & 0.215 & 0.431 & 0.559 & 0.218 & 0.281 \\[-0.1 cm] 
   \hline
\end{tabular}
\end{table}

\newpage
\clearpage

\begin{table}[ht]
	\centering
	\caption{ MC Means and standard deviations of estimators of fixed model parameters. True values are $(\beta_{0}, \beta_{1}, \sigma^{2}_{b}) = (1, 2, 0.36)$. $t_{(3)}$ distributions; unequal $\bm{\Psi}_{i}$. } \label{tabt31} 
		\setlength{\tabcolsep}{5pt} 
	\renewcommand{\arraystretch}{1.5}
	\begin{tabular}{ccccccccccc}
		\hline
	   &   & & & &\multicolumn{2}{c}{\underline{ME-Cor}} &\multicolumn{2}{c}{\underline{YL}} & \multicolumn{2}{c}{\underline{FH}} \\[-0.1 cm]
	 & $a$ & $b$ &$\rho$& $n$ & MC Mean & MC SD & MC Mean & MC SD & MC Mean & MC SD \\[-0.1 cm] 
		\hline
$\beta_{0}$ &   &   &   & 100  & 0.999 & 0.314 & 0.976 & 0.277 & 1.206 & 0.280 \\[-0.1 cm] 
$\beta_{1}$ & 0.250 & 0.750 & 0.200 & 100 & 2.003 & 0.054 & 2.007 & 0.048 & 1.962 & 0.048 \\[-0.1 cm] 
$\sigma^{2}_{b}$ &   &   &   & 100 & 0.402 & 1.697 & 0.252 & 1.765 & 1.022 & 1.789 \\[-0.1 cm] 
\hline
$\beta_{0}$ &   &   &   & 500 & 0.984 & 0.186 & 0.949 & 0.169 & 1.233 & 0.168 \\[-0.1 cm] 
$\beta_{1}$ & 0.250 & 0.750 & 0.200 & 500 & 2.003 & 0.034 & 2.010 & 0.030 & 1.954 & 0.030 \\[-0.1 cm] 
$\sigma^{2}_{b}$ &   &   &   & 500 & 0.348 & 0.569 & 0.135 & 0.549 & 1.012 & 0.608 \\[-0.1 cm] 
\hline
$\beta_{0}$ &   &  &   & 100 & 1.000 & 0.241 & 0.811 & 0.236 & 1.082 & 0.227 \\[-0.1 cm] 
$\beta_{1}$ & 0.250 & 0.750 & 0.800 & 100 & 2.000 & 0.046 & 2.039 & 0.044 & 1.983 & 0.042 \\[-0.1 cm] 
$\sigma^{2}_{b}$ &   &   &   & 100 & 0.355 & 0.592 & 0.036 & 0.465 & 0.114 & 0.525 \\[-0.1 cm] 
\hline
$\beta_{0}$ &  &   &   & 500 & 0.998 & 0.105 & 0.804 & 0.097 & 1.086 & 0.093 \\[-0.1 cm] 
$\beta_{1}$ & 0.250 & 0.750 & 0.800 & 500 & 2.000 & 0.020 & 2.041 & 0.018 & 1.982 & 0.017 \\[-0.1 cm] 
$\sigma^{2}_{b}$ &   &   &  & 500 & 0.361 & 0.548 & 0.020 & 0.441 & 0.078 & 0.538 \\[-0.1 cm] 
\hline
$\beta_{0}$ &   &  &   & 100 & 0.898 & 0.570 & 0.897 & 0.508 & 1.810 & 0.456 \\[-0.1 cm] 
$\beta_{1}$ & 0.750 & 0.250 & 0.200 & 100 & 2.024 & 0.117 & 2.021 & 0.100 & 1.824 & 0.092 \\[-0.1 cm] 
$\sigma^{2}_{b}$ &   &   &   & 100 & 0.505 & 1.101 & 0.450 & 1.184 & 3.225 & 1.388 \\[-0.1 cm] 
\hline
$\beta_{0}$ &   &   &   & 500 & 0.941 & 0.345 & 0.923 & 0.294 & 1.869 & 0.278 \\[-0.1 cm] 
$\beta_{1}$ & 0.750 & 0.250 & 0.200 & 500 & 2.012 & 0.070 & 2.016 & 0.060 & 1.819 & 0.056 \\[-0.1 cm] 
$\sigma^{2}_{b}$ &   &   &   & 500 & 0.385 & 0.780 & 0.265 & 0.834 & 3.317 & 0.899 \\[-0.1 cm] 
\hline
$\beta_{0}$ &   &   &   & 100 & 0.959 & 0.447 & 0.845 & 0.389 & 1.528 & 0.384 \\[-0.1 cm] 
$\beta_{1}$ & 0.750 & 0.250 & 0.800 & 100 & 2.007 & 0.080 & 2.031 & 0.070 & 1.898 & 0.068 \\[-0.1 cm] 
$\sigma^{2}_{b}$ &   &   &   & 100 & 0.456 & 1.234 & 0.169 & 1.058 & 2.168 & 1.348 \\[-0.1 cm] 
\hline
$\beta_{0}$ &   &   &   & 500 & 0.966 & 0.325 & 0.791 & 0.304 & 1.696 & 0.270 \\[-0.1 cm] 
$\beta_{1}$ & 0.750 & 0.250 & 0.800 & 500 & 2.007 & 0.066 & 2.045 & 0.063 & 1.854 & 0.054 \\[-0.1 cm] 
$\sigma^{2}_{b}$ &   &   &  & 500 & 0.384 & 0.817 & 0.069 & 0.622 & 2.127 & 0.809 \\[-0.1 cm] 
   \hline
\end{tabular}
\end{table}

\newpage
\clearpage

\begin{table}[ht]
	\centering
	\caption{ MC Means and standard deviations of estimators of fixed model parameters. True values are $(\beta_{0}, \beta_{1}, \sigma^{2}_{b}) = (1, 2, 0.36)$. $\chi^{2}_{(3)}$ distributions; unequal $\bm{\Psi}_{i}$.} \label{tabchi1} 
	\setlength{\tabcolsep}{5pt} 
	\renewcommand{\arraystretch}{1.5}	
	\begin{tabular}{ccccccccccc}
		\hline
	    &  & & & &\multicolumn{2}{c}{\underline{ME-Cor}} &\multicolumn{2}{c}{\underline{YL}} & \multicolumn{2}{c}{\underline{FH}} \\[-0.1 cm]
	 & $a$ & $b$ &$\rho$& $n$ & MC Mean & MC SD & MC Mean & MC SD & MC Mean & MC SD \\[-0.1 cm] 
		\hline
$\beta_{0}$ &   &  &  & 100  & 1.002 & 0.287 & 0.954 & 0.273 & 1.243 & 0.264 \\[-0.1 cm]  
$\beta_{1}$ & 0.250 & 0.750 & 0.200 & 100  & 2.000 & 0.046 & 2.007 & 0.042 & 1.958 & 0.041 \\[-0.1 cm]  
$\sigma^{2}_{b}$ &   &   &   & 100  & 0.347 & 0.298 & 0.123 & 0.259 & 1.043 & 0.382 \\[-0.1 cm]  
\hline 
$\beta_{0}$ &   &   &  & 500 & 0.993 & 0.129 & 0.946 & 0.120 & 1.254 & 0.116 \\[-0.1 cm]  
$\beta_{1}$ & 0.250 & 0.750 & 0.200 & 500 & 2.000 & 0.025 & 2.011 & 0.023 & 1.946 & 0.022 \\[-0.1 cm]  
$\sigma^{2}_{b}$ &   &   &  & 500 & 0.358 & 0.146 & 0.043 & 0.091 & 1.044 & 0.170 \\[-0.1 cm]  
\hline 
$\beta_{0}$  &   &   &  & 100 & 1.002 & 0.164 & 0.857 & 0.167 & 1.063 & 0.161 \\[-0.1 cm]  
$\beta_{1}$ & 0.250 & 0.750 & 0.800 & 100 & 2.000 & 0.028 & 2.027 & 0.027 & 1.988 & 0.026 \\[-0.1 cm]  
$\sigma^{2}_{b}$ &   &   &   & 100 & 0.351 & 0.145 & 0.000 & 0.000 & 0.071 & 0.101 \\[-0.1 cm]  
\hline 
$\beta_{0}$  &   &   &   & 500 & 0.998 & 0.071 & 0.846 & 0.072 & 1.066 & 0.070 \\[-0.1 cm]  
$\beta_{1}$ & 0.250 & 0.750 & 0.800 & 500 & 2.000 & 0.012 & 2.029 & 0.012 & 1.987 & 0.011 \\[-0.1 cm]  
$\sigma^{2}_{b}$ &   &   &   & 500 & 0.355 & 0.065 & 0.000 & 0.000 & 0.045 & 0.048 \\[-0.1 cm] 
\hline 
$\beta_{0}$  &   &   &   & 100 & 0.960 & 0.462 & 0.912 & 0.400 & 2.022 & 0.332 \\[-0.1 cm] 
$\beta_{1}$ & 0.750 & 0.250 & 0.200 & 100 & 2.010 & 0.112 & 2.021 & 0.099 & 1.782 & 0.082 \\[-0.1 cm] 
$\sigma^{2}_{b}$ &   &   &   & 100 & 0.457 & 0.573 & 0.332 & 0.635 & 3.382 & 0.823 \\[-0.1 cm] 
\hline 
$\beta_{0}$  &   &   &   & 500 & 0.999 & 0.171 & 0.952 & 0.152 & 1.857 & 0.134 \\[-0.1 cm] 
$\beta_{1}$ & 0.750 & 0.250 & 0.200 & 500 & 2.001 & 0.038 & 2.010 & 0.033 & 1.827 & 0.029 \\[-0.1 cm] 
$\sigma^{2}_{b}$ &   &   &   & 500 & 0.368 & 0.287 & 0.138 & 0.260 & 3.471 & 0.406 \\[-0.1 cm] 
\hline 
$\beta_{0}$  &   &   &   & 100 & 0.994 & 0.320 & 0.830 & 0.290 & 1.630 & 0.258 \\[-0.1 cm] 
$\beta_{1}$ & 0.750 & 0.250 & 0.800 & 100 & 2.000 & 0.064 & 2.032 & 0.058 & 1.879 & 0.051 \\[-0.1 cm] 
$\sigma^{2}_{b}$ &   &  &   & 100 & 0.414 & 0.442 & 0.025 & 0.184 & 2.272 & 0.637 \\[-0.1 cm] 
\hline 
$\beta_{0}$  &   &   &   & 500 & 0.997 & 0.135 & 0.848 & 0.123 & 1.603 & 0.111 \\[-0.1 cm] 
$\beta_{1}$ & 0.750 & 0.250 & 0.800 & 500 & 2.001 & 0.027 & 2.029 & 0.023 & 1.884 & 0.022 \\[-0.1 cm] 
$\sigma^{2}_{b}$ &   &   &  & 500 & 0.354 & 0.194 & 0.000 & 0.008 & 2.236 & 0.264 \\[-0.1 cm] 
   \hline
\end{tabular}
\end{table}

\newpage
\clearpage

\begin{table}[ht]
\centering 
	\caption{ MC MSPE's of alternative predictors (MC MSPE of Predictor), and MC means of estimated MSPE's of ME-Cor and FH predictors (MC Mean Est. MSPE).  $\chi^{2}_{(3)}$ distributions; unequal $\bm{\Psi}_{i}$. }\label{tabchi2} 
	\setlength{\tabcolsep}{5pt} 
	\renewcommand{\arraystretch}{1.5}	
	\begin{tabular}{cccccccccc}
		\hline
		& & & & \multicolumn{ 4}{c}{\underline{MC MSPE of Predictor}} & \multicolumn{2}{c}{\underline{MC Mean Est. MSPE}}  \\
		\hline \hline 
	\multicolumn{3}{c}{$(a, b, \rho)$} & $n$ & Direct &   ME-Cor&  YL & FH & ME-Cor-MSPE & FH-MSPE \\[-0.1 cm] 
		\hline
 \multicolumn{3}{c}{(0.250, 0.750, 0.200)}& 100  & 1.012 & 0.755 & 0.764 & 0.828 & 0.743 & 0.492 \\[-0.1 cm]  
    &   &   & 500 & 1.007 & 0.743 & 0.759 & 0.807 & 0.740 & 0.485 \\[-0.1 cm]   
  \hline 
  \multicolumn{3}{c}{(0.250, 0.750, 0.800)} & 100  & 1.016 & 1.009 & 1.116 & 1.589 & 1.001 & 0.094 \\[-0.1 cm]   
    &   &   & 500  & 1.006 & 0.999 & 1.108 & 1.615 & 1.001 & 0.048 \\[-0.1 cm]   
  \hline 
  \multicolumn{3}{c}{(0.750, 0.250, 0.200)} & 100  & 0.336 & 0.334 & 0.343 & 0.355 & 0.334 & 0.301 \\[-0.1 cm]   
    &   &   & 500 & 0.335 & 0.333 & 0.344 & 0.355 & 0.333 & 0.301 \\[-0.1 cm]   
  \hline 
\multicolumn{3}{c}{(0.750, 0.250, 0.800)} & 100  & 0.337 & 0.215 & 0.444 & 0.563 & 0.218 & 0.286 \\[-0.1 cm]   
    &   &   & 500 & 0.336 & 0.213 & 0.444 & 0.565 & 0.212 & 0.285 \\[-0.1 cm]   
   \hline
\end{tabular}
\end{table}

\subsection{Bivariate covariate} 

We validate the procedure for a model with two covariates, both of which are subject to measurement error. The simulation model is defined by 
\begin{align*}
    Y_{i} = 1 + 2x_{1i} + 3x_{2i} + b_{i} + e_{i},
\end{align*}
where $b_{i}\stackrel{iid}{\sim}N(0, 0.36)$ and the distribution of $e_{i}$ is defined below. Instead of observing $x_{1i}$ and $x_{2i}$, we observe 
\begin{align*}
    (W_{1i}, W_{2i})' = (x_{1i}, x_{2i})'+ (u_{1i}, u_{2i})',
\end{align*}
where 
\begin{align*}
(e_{i}, u_{1i},  u_{2i})' \stackrel{ind}{\sim} N_{3}(\bm{0}, \bm{\Psi}_{i}),  
\end{align*}
and
\begin{align*}
\bm{\Psi}_{i} = \left( \begin{array}{ccc} 0.250 & 0.071 & 0.086 \\
                                    0.071 & 0.500 & 0.122 \\
                                    0.086 & 0.122 & 0.740\end{array}\right)
\end{align*}
for all $i= 1,\ldots, n$. As the model has more parameters than the model with a univariate covariate, we set $n = 2000$ for this simulation. 

We calculate the estimators of fixed parameters, predictors, and estimator of the MSPE according to the proposed procedures. The MC mean of the estimators of $(\beta_{0}, \beta_{1}, \beta_{2})' = (1, 2, 3)'$ is $(0.9904, 2.0003, 3.0017)'$. The MC mean of the estimator of $\sigma^{2}_{b} = 0.36$ is 0.349. The average MC MSPE of the predictors is 0.2479. The average of the estimated MSPE is 0.2478. This validates the properties of the proposed procedure for the case of a bivariate covariate.

\section{Comparison to \cite{burgard2022measurement}}

We compare the proposed (ME-Cor) predictor to the predictor derived using the method of \cite{burgard2022measurement}. As in \cite{burgard2022measurement}, we use maximum likelihood to estimate the fixed model parameters. We denote the maximum likelihood estimators by $(\hat{\beta}_{0,ML}, \hat{\beta}_{1,ML}, \hat{\sigma}^{2}_{b,ML})'$. We define a predictor based on the method of  \cite{burgard2022measurement} for a univariate version of the measurement error model. We define the predictor of  \cite{burgard2022measurement} as 
\begin{align*}
\hat{\theta}_{i,bg} = \hat{\beta}_{0,ML} + \hat{\beta}_{1,ML}W_{i} + \hat{\beta}_{1,ML}\hat{v}_{i,bg} + \hat{b}_{i,bg},
\end{align*}
where 
\begin{align*}
\hat{v}_{i,bg}  & = R_{i}(Y_{i} - \hat{\beta}_{0,ML} - \hat{\beta}_{1,ML}W_{i}), \\ \nonumber
R_{i} & = R_{i,n}R_{i,d}^{-1}, \\ \nonumber 
R_{i,n} &=  (\hat{\beta}_{1,ML} + \psi_{uei}/\psi^{2}_{ui})/(\hat{\sigma}^{2}_{b,ML} + \psi^{2}_{ei} - \psi^{2}_{uei}/\psi^{2}_{ui}) ,\\ \nonumber 
R_{i,d}& = (\hat{\beta}_{1,ML} + \psi_{uei}/\psi^{2}_{ui})^{2}/(\hat{\sigma}^{2}_{b,ML} + \psi^{2}_{ei} - \psi^{2}_{uei}/\psi^{2}_{ui}) + \psi^{-2}_{ui},  \\ \nonumber
\hat{b}_{i,bg} & =  \frac{ \hat{\sigma}^{2}_{b,ML}  }{ \hat{\sigma}^{2}_{b,ML} + \hat{\sigma}^{2}_{\delta i,ML}   },
\end{align*}
and $\hat{\sigma}^{2}_{\delta i,ML} =  \hat{\beta}_{1,ML}^{2}\psi^{2}_{ui} + 2\hat{\beta}_{1,ML}\psi_{ue i} + \psi^{2}_{ei}$.
For the simulation configuration in Table~\ref{tabbg}, the proposed predictor has smaller average MSE than the predictor of  \cite{burgard2022measurement}. 

\begin{table}[ht]
\centering
\caption{Average MSEs of proposed predictor (ME-Cor) and  \cite{burgard2022measurement} predictor. }\label{tabbg}
	\setlength{\tabcolsep}{5pt} 
	\renewcommand{\arraystretch}{1.5}
\begin{tabular}{cccc}
  \hline
$(a,b,\rho)$ & $n$   & Burgard & ME-Cor \\ 
  \hline
(0.25, 0.75, 0.20) &100 & 0.743 & 0.732 \\ 
(0.25, 0.75, 0.20) &500 & 0.760 & 0.747 \\ 
(0.25, 0.75, 0.80) &100 & 1.115 & 1.006 \\ 
(0.25, 0.75, 0.80) &500 & 1.087 & 0.998 \\ 
(0.75, 0.25, 0.20) &100 & 0.377 & 0.342 \\ 
(0.75, 0.25, 0.20) &500 & 0.370 & 0.335 \\ 
(0.75, 0.25, 0.80) &100 & 0.681 & 0.215 \\ 
(0.75, 0.25, 0.80) &500 & 0.695 & 0.214 \\ 
  \hline
\end{tabular}
\end{table}

\end{document}